\documentclass[12pt]{article}
\usepackage[margin=1.2in]{geometry}
\usepackage{amsmath}
\usepackage{amssymb}
\usepackage{relsize}
\usepackage{bigints}
\usepackage{natbib}
\usepackage{graphicx}
\usepackage{xcolor}
\usepackage{bm}
\usepackage[labelfont={bf,small}, textfont=small, skip=1pt, font={stretch=1.2}]{caption}
\usepackage{upgreek}
\usepackage[utf8]{inputenc}
\usepackage{nameref}
\usepackage[colorlinks=true, allcolors=blue]{hyperref}
\usepackage{cleveref}
\usepackage{xfrac}
\usepackage{longtable}
\usepackage{setspace}
\usepackage{todonotes}

\def\correspondingauthor{\footnote{To whom correspondence must be addressed: sifraley@ucsd.edu and prangamani@ucsd.edu}}

\newcommand{\beginsupplement}{%
        \setcounter{table}{0}
        \renewcommand{\thetable}{S\arabic{table}}%
        \setcounter{figure}{0}
        \renewcommand{\thefigure}{S\arabic{figure}}%
        \setcounter{section}{0}
        \renewcommand{\thesection}{S\arabic{section}}
        \setcounter{subsection}{0}
        \renewcommand{\thesubsection}{SI\arabic{subsection}}
     }

\usepackage{mathtools,tikz,caption}
\usepackage{tikz}
\usepackage{authblk}
\title{Modeling collagen fibril degradation as a function of matrix microarchitecture}
\author[1]{B. \textbf{Debnath}}
\author[2]{B. N. \textbf{Narasimhan}}
\author[*2]{S. I. \textbf{Fraley}}
\author[1,3]{P. \textbf{Rangamani} \correspondingauthor{}}
\affil[1]{Department of Mechanical and Aerospace Engineering, University of California San Diego, CA 92093, USA}
\affil[2]{Department of Bioengineering, University of California San Diego, CA 92093, USA}
\affil[3]{Department of Pharmacology, School of Medicine, University of California San Diego, CA 92093, USA}

\usepackage{mathtools}
\usepackage{tikz-cd}
\usepackage{bm}
\usepackage{epstopdf}
\AppendGraphicsExtensions{.tif}
\def\md{\mbox{d}}
\date{}     

\begin{document}
    
\maketitle

\begin{abstract}

\noindent
Collagenolytic degradation is a process fundamental to tissue remodeling.
The microarchitecture of collagen fibril networks changes during development, aging, and disease.
Such changes to microarchitecture are often accompanied by changes in matrix degradability. 
{\it In vitro}, collagen matrices of the same concentration but different microarchitectures also vary in degradation rate.   
How do different microarchitectures affect matrix degradation? 
To answer this question, we developed a computational model of collagen degradation. 
We first developed a lattice model that describes collagen degradation at the scale of a single fibril. 
We then extended this model to investigate the role of microarchitecture using Brownian dynamics simulation of enzymes in a multi-fibril three dimensional matrix to predict its degradability. 
Our simulations predict that the distribution of enzymes around the fibrils is non-uniform and depends on the microarchitecture of the matrix. This non-uniformity in enzyme distribution can lead to different extents of degradability for matrices of different microarchitectures.
Our model predictions were tested using {\it in vitro} experiments with synthesized collagen gels of different microarchitectures.
Experiments showed that indeed degradation of collagen depends on the matrix architecture and fibril thickness.
In summary, our study shows that the microarchitecture of the collagen matrix is an important determinant of its degradability. 

\end{abstract}

\section{Introduction}

Collagen is the most abundant protein present in tissues. Enzymatic degradation of collagen is an important process in both physiological and pathological conditions \citep{wohlgemuth2023alignment,
gupta20243d}. For instance, an imbalance between collagen degradation and production can lead to increased collagen accumulation resulting in fibrosis  \citep{wynn2012mechanisms, mckleroy2013always}.
Such fibrotic environments have been associated with an impaired degradative environment \citep{mckleroy2013always}. Studies have shown that the microarchitecture of collagen in the fibrotic extracellular matrix (ECM) is substantially different from the healthy tissues and they show higher resistance to degradation 
\citep{huang2009transglutaminase, philp2018extracellular}.
As another example, cancer-associated fibroblasts in the tumor microenvironment secrete collagen and continuously remodel their matrix. 
This abnormal remodeling can lead to the cancerous microenvironment possessing a higher density of collagen and very different microarchitecture than a healthy ECM \citep{nebuloni2016insight}. 
A body of work argues that  alterations in collagen degradation can result in metastasis  \citep{liotta1980metastatic, winkler2020concepts}, and suggests possible connections between the matrix microarchitecture and its degradability \citep{dewavrin2014tuning,
ranamukhaarachchi2019macromolecular,
ashworth2024importance, narasimhan2024degradability}. 
However, the role of matrix microarchitecture in determining the degradability at the cellular length scale remains an open question. 
Here, we used multi-scale modeling and \textit{in vitro} experiments to investigate possible mechanisms of matrix degradation at this length scale.
 
Most models of collagen gel degradation use Michaelis-Menten kinetics to model the collagenase activity and show good agreement with experiments in predicting the rate of total mass loss \citep{tzafriri2002reaction, metzmacher2007model, ray2013drug, vuong2017biochemo}. 
These models cannot address the connection between the microarchitecture and  degradation because of their continuum structure. 
Another class of models have treated collagen fibril as a thin filament of  negligible thickness (diameter $\sim 10-20$ nm) and proposed a degradation mechanism of one filament based on movements of a few collagenase molecules over the filament surface \citep{saffarian2004interstitial, sarkar2012single}.
However, many experimental findings have reported that the collagen fibrils in a matrix are significantly thicker (diameter in the range $\sim 0.1-0.5 \, \mu$m) than the value used in these models \citep{bhole2009mechanical, flynn2013highly, staunton2016mechanical, ranamukhaarachchi2019macromolecular, seo2020collagen}. 
Therefore, an open challenge in the field is to determine  how a filament-scale model can be extended to a fibril-scale model, and subsequently, to a model where multiple fibrils are interacting with many collagenase molecules in a three-dimensional matrix.

In this work, we investigated how matrix microarchitecture can affect its degradability. We  developed a fibril-scale model using lattice-based approach to predict the degradation of single fibril. 
To predict degradation of multiple fibrils by many enzyme molecules in a matrix, we used Brownian dynamics to capture the enzyme distribution surrounding the fibrils in a three-dimensional matrix environment.
We then combined  the fibril-scale model with the enzyme distribution obtained from the Brownian dynamics simulations to predict the degradation of collagen matrices. Our model predicted that differences in the microarchitecture between two matrices of same collagen density can lead to different extents of degradation.  Additionally, we predicted that fibril thickness can be an important determinant of degradation.
We tested this prediction against {\textit{in vitro}} experiments using collagen gels of different architectures, which showed that indeed, fibril degradation depends on the matrix microarchitecture. 
Thus our study may provide new insights for understanding matrix alterations associated with disease and  may influence the development of matrix targeted therapeutics, biomaterials and controlled drug delivery \citep{daly2020hydrogel, kim2022tissue, gupta20243d}.

\section{Methods}
In this section, we describe the details of the single fibril model, Brownian dynamics (BD) simulations, and experimental methods.
We elaborate on each step of the model development and justify the assumptions based on previous experimental observations.
The notation and  symbols used in this work are shown in Table~\ref{tab:table_symbols}.

\subsection{Model development for the degradation of single fibril}
\label{sec:model_dev}

\begin{figure}
\centering
\includegraphics[width=1\linewidth]{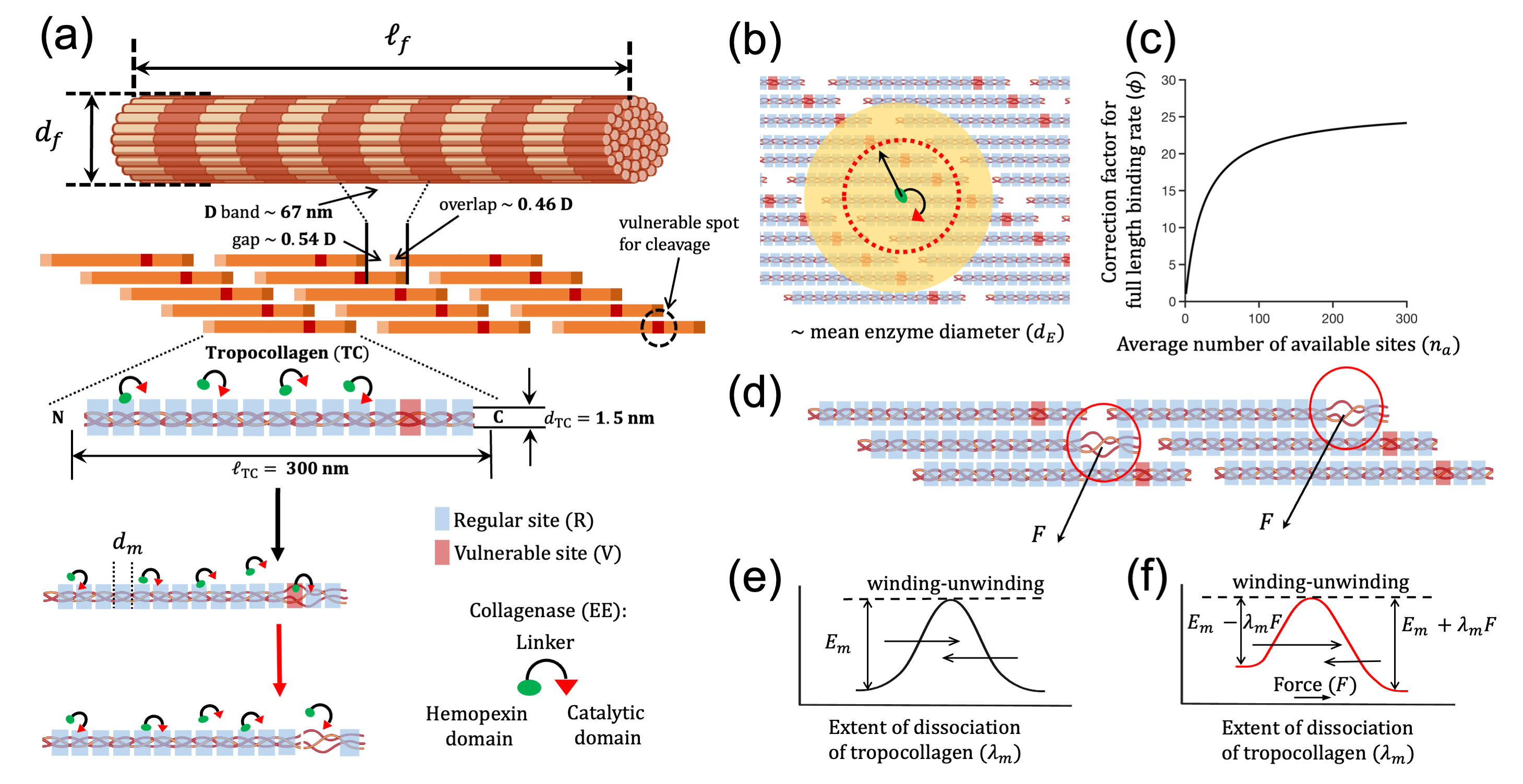}
\caption{{\textbf{Modeling  single fibril degradation.}} (a) Overview of the  hierarchical organization of a collagen fibril and the degradation mechanism. We divide  the tropocollagen unit into lattice sites (regular and vulnerable sites). The structure of the collagenase enzymes has two domains. (b) When one domain of the enzyme is bound to one lattice site, the other domain has access to only the neighboring unoccupied sites inside the red-dashed circle, whose radius is equivalent to the hydrodynamic mean diameter ($d_E$) of the enzyme. (c) To incorporate this effect in the intrinsic rates  of the full-length binding (when both domains of an enzyme are bound to the lattice sites)/hopping kinetics, we propose a correction factor ($\phi$)  as a function of available sites ($n_a$). (d) We propose that the enzyme-induced irreversible unwinding of the vulnerable sites induces a force, which affects the unwinding rate of the neighboring regular sites. The symmetric (e) and asymmetric (f) energy barriers  for the winding-unwinding process of a lattice site. }
\label{fig:schematic_model}
\end{figure}

We consider a single collagen fibril of mean diameter $d_f$ and mean length $\ell_f$.
This fibril consists of a staggered arrangement of the triple helix tropocollagen  units as shown in Fig.~\ref{fig:schematic_model}a \citep{orgel2006microfibrillar, buehler2008nanomechanics, marino2014stress, saini2020tension}. 
The enzymatic degradation of the fibril occurs in the presence of collagenases such as matrix metalloproteinases (MMP). 
These collagenases cleave the triple helix at a site that is at a distance $\sim$ 67 nm (approx. $1/4$ of the length of the triple helix)  from the C-terminal of the tropocollagen unit (Fig.~\ref{fig:schematic_model}a) \citep{chung2004collagenase, nagase2011triple}.

Previous experiments have reported that the 
collagenolytic degradation is a surface erosion process \citep{okada1992degradation, perumal2008collagen, tzafriri2002reaction}. 
The minimum  effective pore size inside a collagen fibril is a few multiples of the diameter of the tropocollagen, which is $ d_{\text{TC}} \sim  1.5$ nm \citep{orgel2006microfibrillar}. 
This dimension is smaller than the size of the collagenase molecules which is in the range 50-120 kDa with a mean hydrodynamic diameter of $ d_E \sim  10-20$ nm \citep{tyn1990prediction}.
Based on the dimensions and following previous studies \citep{okada1992degradation, perumal2008collagen, tzafriri2002reaction}, we assume that the diffusion of MMPs through the pores of the fibril can be ignored, and we model the fibril degradation as a surface erosion process.

To simulate the surface erosion of the fibril, we developed a lattice-based model.
We divided the tropocollagen unit into lattice sites each of length $d_m$. 
Based on the number of amino acid residues at the cleavage site region \citep{perumal2008collagen}, we set $d_m = 8$ nm.
As the cleavage site is at a distance $1/4$ of the length of the tropocollagen from its C-terminal, we assumed that there is one vulnerable site (V) per tropocollagen unit, and we treat the rest of the lattice sites  as regular sites (R) (see Fig.~\ref{fig:schematic_model}a). With these dimensions, we developed a relation between the size of a fibril and the number of lattice sites available at its surface.
To obtain the total number of lattice sites on the fibril surface, we next estimate the number of tropocollagen units on that surface. At any given time $t$, the number of tropocollagen units available for collagenolysis at the fibril surface  is
\begin{equation}
    N^s_{\text{TC}} = \frac{\pi \, d_f \, \ell_f}{d_{\text{TC}} \, (\ell_{\text{TC}} + 0.54\, {\text D})},
    \label{eq:tc_units}
\end{equation}
where  $\ell_{\text{TC}}$ is the  length of the tropocollagen unit and (0.54 D) is the gap between two tropocollagen units (D  $\sim$ 67 nm) (see  Fig.~\ref{fig:schematic_model}a). Thus, the  total number of sites $N^s$ at time $t$ exposed at the fibril surface is  
\begin{equation}
    N^s = \frac{\ell_{\text{TC}}}{d_m} \, N^s_{\text{TC}} = \frac{\ell_{\text{TC}}}{d_m} \, \frac{\pi \, d_f \, \ell_f}{d_{\text{TC}} \, (\ell_{\text{TC}} + 0.54\, {\text D})}.
    \label{eq:total_sites}
\end{equation}
Eqn (\ref{eq:total_sites}) highlights the relationship between the total number of sites available for degradation and the size of the fibril. The number of vulnerable sites $N^s_V$ exposed at the surface is the same as $N^s_{\text{TC}}$ because one tropocollagen unit contains one vulnerable site, i.e., $N^s_V = N^s_{\text{TC}}$. Therefore the remaining $N^s_R = (N^s - N^s_V)$ are  regular sites. Using eqn (\ref{eq:total_sites}), we can estimate either $\ell_f$ or $d_f$. As the vulnerable sites are not aligned at the same plane inside the fibril (see Fig.~\ref{fig:schematic_model}a) and the collagenolytic degradation is a surface erosion process, we treat  $\ell_f$ as a constant and predict  $d_f$ from eqn (\ref{eq:total_sites}).

\begin{table}
    \centering
        \caption{Notation and list of symbols.}
    \begin{tabular}{l l}
         \hline 
Notation &  Description \\
\hline
$d_f$ & mean diameter of a fibril \\
$\ell_f$ & mean length of a fibril \\
$d_{\text{TC}}$ & diameter of the tropocollagen unit  \\
$\ell_{\text{TC}}$ & length of the tropocollagen unit \\
$d_E$ & mean hydrodynamic diameter of an enzyme \\
$d_m$ & length of one lattice site \\
R & regular site \\
V & vulnerable site \\
$EE$ & enzyme with two domains  \\
 $E_{*} E_{*}$ class of symbols & enzyme partially ($E_{*} E$) or, fully  bound ($E_{*} E_{*}$) \\
 & to regular ($* = R$) or, vulnerable ($* = V$) sites\\
 $k^{*}_*$ class of symbols & different  rate constants \\ 
$N^s$ & total sites exposed on the fibril surface\\
$N^s_R$ & number of regular sites exposed on the fibril surface\\
$N^s_V$ & number of vulnerable sites exposed on the fibril surface\\
$N_{EE}$ & number of free enzymes \\
 $N^s_{E_* E_*}$ class of symbols & number of  enzymes partially (or, fully) bound \\
 & to regular (or, vulnerable) sites\\
$\phi$ & correction factor for intrinsic rates of  full-length binding kinetics \\
$n_a$ & number of available lattice sites per enzymes at partially-bound state \\
 $E_m$ & energy required to cross the symmetric energy barrier \\
 $\lambda_m$ & extent of dissociation at the unwound state \\
 $F$ & average force experienced by the lattice sites  \\
 $\gamma$ & surface energy per unit area \\
 $\tau_c$ & characteristic time associated with force dependent kinetics\\
 $\tau_e$ & time for disentanglement of tropocollagen chains \\
 $l_p$ & nominal length scale \\
  $N_e^0$ & number of enzymes surrounding a fibril \\
$(N_e^0)_{total}$ & total number of enzymes in the simulation box \\
 $D_e$ & diffusivity of enzymes \\
 $\phi_f$ & volume fraction of fibrils \\
 $n_f$ & number of fibrils in the simulation box \\
 $A_f^0$ & initial surface area of a fibril \\
 $(A_f^0)_{total}$ & initial total surface area of all fibrils \\
\hline
    \end{tabular}
\label{tab:table_symbols}
\end{table}

\subsubsection*{Development of a reaction scheme to model the loss of lattice sites on a fibril}

\begin{figure}[hbtp]
\includegraphics[width=1\linewidth]{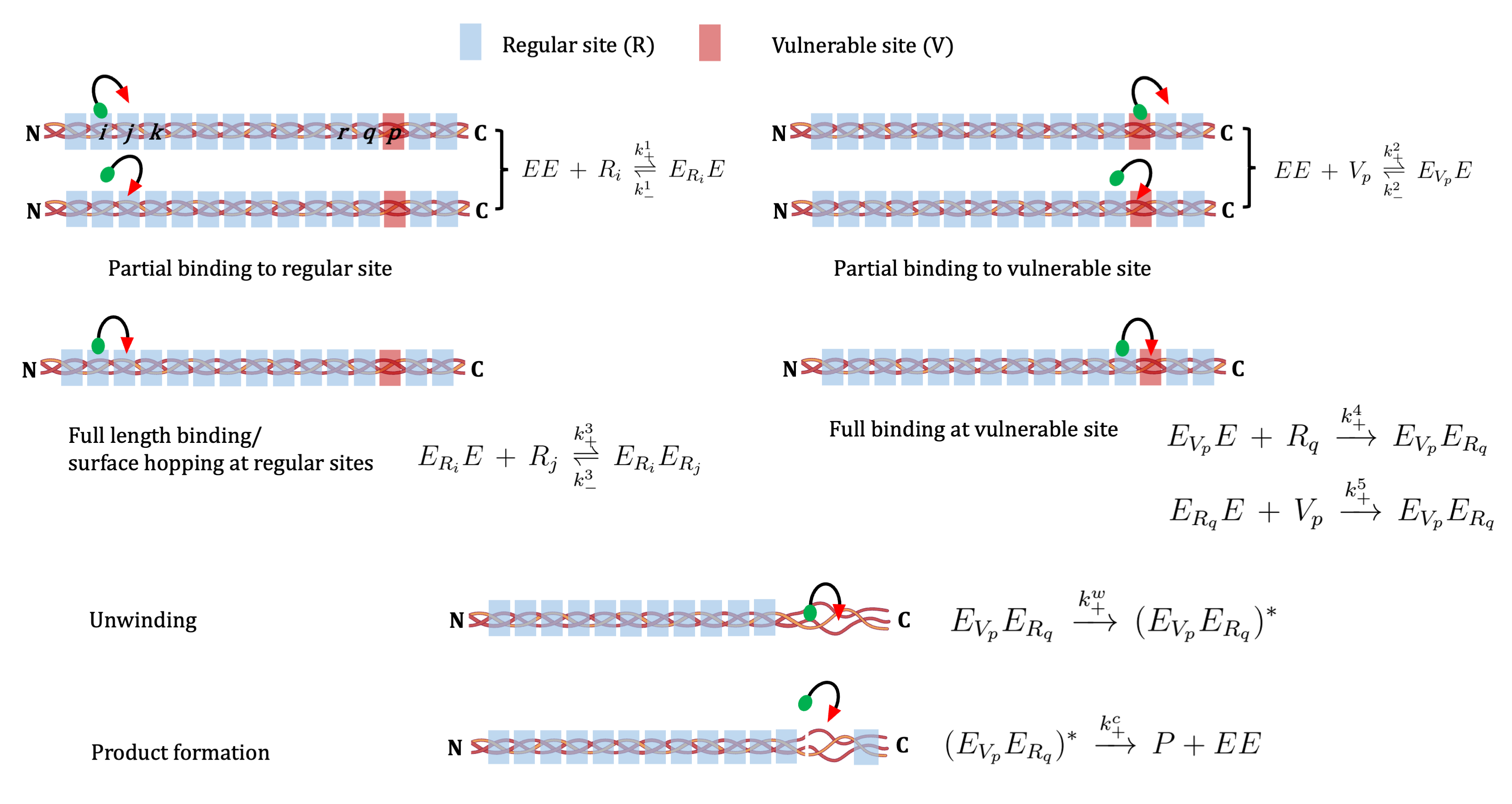}
\caption{Schematic of the enzyme kinetics.}
 \label{fig:sup1_kinetics_schematic}
\end{figure}

Previous experimental findings have shown that the MMP class proteinases (for example, MMP1) hydrolyze the peptide bonds at the vulnerable site  of the collagen type I  \citep{chung2004collagenase, nagase2011triple}. 
Researchers hypothesized that the enzyme destabilizes the structure of the tropocollagen, and induces local unwinding around the vulnerable site before cleavage \citep{chung2004collagenase,han2010molecular,perumal2008collagen}.
Using the hypotheses of \cite{chung2004collagenase} and \cite{perumal2008collagen}, and making the following assumptions,  we propose a reaction scheme for enzyme kinetics (see Fig.~\ref{fig:schematic_model}a and the supplementary information (SI) section SI1 (Fig.~\ref{fig:sup1_kinetics_schematic})):

\begin{enumerate}

\item There are two types of lattice sites on the collagen surface: regular site (R) and vulnerable site (V). The MMP has two domains: Hemopexin (Hpx) domain and Catalytic (Cat) domain connected by a linker \citep{visse2003matrix, iyer2006crystal}. 
\cite{manka2012structural} had shown a strict requirement of both domains to be linked together for efficient enzyme-collagen binding and collagenolysis. 

\item Following \cite{manka2012structural}, we assume that the collagenase molecule has two domains, denoted by $EE$. Using these domains, the enzyme binds to the lattice sites via adsorption-desorption mechanism and hops on the sites. For the sake of simplicity, we assume that these two domains are equivalent and are of the same size.   

\item First, using any one domain, the enzyme binds to a single site, in what is denoted as `reversible partial binding'. This type of binding can result in two possible reactions. 

\begin{equation}
EE \, + \, R \, \underset{k^1_{-}}{\stackrel{k^1_{+}}{\rightleftharpoons}} \, E_R E  \hspace{3cm}   EE \, + \, V \, \underset{k^2_{-}}{\stackrel{k^2_{+}}{\rightleftharpoons}} \, E_V E
\label{eq:kinetics1}
\end{equation}

For the reversible partial binding (eqn (\ref{eq:kinetics1})), we implement the protein adsorption-desorption kinetics \citep{adamczyk1999irreversible, fang2005kinetics} (see a short note on adsorption kinetics in section SI2). 

\item When one domain of the enzyme is bound to one site, the other domain can also bind to another site. When both domains of an enzyme are in bound state, we term this state as `full-length binding'.  The enzyme molecule can jump or change track via hopping \citep{saffarian2004interstitial, sarkar2012single}. It hops on the regular sites via reversible binding-unbinding of one domain to a regular site while its other domain remains bound to another regular site.
\begin{equation}
E_R E \, + \, R \, \underset{k^3_{-}}{\stackrel{k^3_{+}}{\rightleftharpoons}} \, E_R E_R  
\label{eq:kinetics2}
\end{equation}

\item Once the enzyme reaches a vulnerable site with both domains in the bound state, hopping stops, and is followed by enzyme-induced unwinding and the formation of product sites.
These product sites  are no longer available for binding new enzyme molecules. 
\begin{equation}
   E_V E \, + \, R \, \xrightarrow{k^4_+} \, E_V E_R  \hspace{3cm} E_R E \, + \, V \, \xrightarrow{k^5_+} \, E_V E_R
   \label{eq:kinetics3}
\end{equation}
\begin{equation}
{\rm{Unwinding:}} \hspace{0.5cm} E_V E_R \,  \xrightarrow{k^w_+} \, (E_V E_R)^*  
\label{eq:kinetics4}
\end{equation}
\begin{equation}
{\rm{Product \, formation:}} \hspace{0.5cm} (E_V E_R)^* \,  \xrightarrow{k^c_+} \, P + EE 
\label{eq:kinetics5}
\end{equation}

\end{enumerate}

For the cases of full-length binding/hopping represented by eqn (\ref{eq:kinetics2})-(\ref{eq:kinetics3}), the forward reactions cannot be treated as second order kinetics  because once one domain of an enzyme is bound to one site, its other domain does not have accessibility to all available lattice sites on the surface (see Fig.~\ref{fig:schematic_model}b). The unbound domain of the enzyme can find another lattice site for  binding  within a searching circle whose radius is equivalent to $d_E$ (the mean hydrodynamic diameter of the enzyme). To correct the intrinsic rates of full-length binding kinetics, we multiply the rates with a correction factor $\phi$.  We propose a phenomenological function for $\phi$ as (Fig.~\ref{fig:schematic_model}c)
\begin{equation}
    \phi  \approx \frac{\pi \, (d_E)^2}{d_m \, d_{\text{TC}}} \; \frac{n_a}{c_1 + n_a},
    \label{eq:correction_phi}
\end{equation}
where $n_a$ is the number of available sites on the fibril surface. See section SI3 for more details of $\phi$ and section SI4 for the system of ODEs representing eqn (\ref{eq:kinetics1})-(\ref{eq:kinetics5}). 

The enzyme-induced permanent unwinding followed by cleavage happens only at the exposed vulnerable sites in collagenolysis. However, it is not clear how the fibril will lose other exposed regular sites at  the surface due to the cleavage events so that the new surfaces can be easily accessible to the enzymes for further degradation \citep{chung2004collagenase,saffarian2004interstitial,nerenberg2008collagenases, perumal2008collagen, sarkar2012single}.  Hence, we propose that the enzyme-induced permanent unwinding at the vulnerable sites generates a force  which assists the removal of more regular sites exposed at the surface. We use the nonequilibrium rate-process theory of Eyring \citep{eyring1935activated,eyring1936viscosity,tobolsky1943mechanical} and propose a new rate-term $\Re$ related to the force-dependent kinetics in an \textit {ad hoc} manner in the system of ODEs. We briefly describe $\Re$ below. 

As a result of thermal fluctuations, the lattice sites  can be  in any state between the  triple helical and  (temporary) unwound configurations  
\citep{perumal2008collagen}. If the rates of transition from the triple helical state to the unwound state and \textit{vice-versa} are equal, there is no net change in the tropocollagen units in absence of enzymes.  We hypothesize that enzyme-induced permanent unwinding that leads to cleavage  generates a local stress (Fig.~\ref{fig:schematic_model}d). This mechanism is similar to the enzyme pulling chew-digest mechanism proposed by \cite{eckhard2011structure}. As a consequence, the other regular sites at the surface  experience an external force which affects the kinetics by increasing their net rate of unwinding 
(Fig.~\ref{fig:schematic_model}e,f). This force can cause the slippage of chains, resulting in detachment from the surface  \citep{adjari1994slippage, sung1995slippage}. 

To incorporate the force-dependent kinetics, we add a new term $\Re = k_R \, N^s_R$  in an \textit{ad hoc} manner to the system of ODEs (see section SI4 for the system of ODEs). Here $k_R$ is  the net rate of flow over the energy barrier towards force assisted unwinding (Fig.~\ref{fig:schematic_model}e,f).  We propose the following expression for $k_R$ \citep{glasstone1941theory, stuart1953dependence}
\begin{equation}
    k_R \, = \, \,  \frac{k_B T}{h} \, \Bigg[{\rm exp} \Bigg(\frac{-(E_m - \lambda_m F)}{k_B T} \Bigg) - {\rm exp} \Bigg(\frac{-(E_m + \lambda_m F)}{k_B T} \Bigg) \Bigg],
\label{eq:rate_theory_kinetic_rate_constant}
\end{equation}
where $h$ is Planck constant, $k_B$ is the Boltzmann constant, and $T$ is the temperature. Here, $E_m$ is the energy required to cross the symmetric barrier (Fig.~\ref{fig:schematic_model}e), $\lambda_m$ is the extent of dissociation in unwound state which is  chosen as $ \sim 3.6 $ \AA $\,$       \citep{perumal2008collagen}. Here $F$ is the average force experienced by the remaining lattice sites. This force changes the energy barrier of unwinding to be asymmetric (Fig.~\ref{fig:schematic_model}f). If $F = 0$, $k_R = 0$. Following a heuristic approach, the average force (per remaining lattice site) is  
\begin{equation}
    F \, \sim \, \frac{1}{N^s} \, \big(\gamma \, d_m \, k^c_{+} \, \tau_{c} \big) \, \, (2 \, k^w_+ \, N^s_{E_V E_R}) \, \tau_e, \label{eq:rate_theory_force}
\end{equation}
where $\gamma \sim \frac{k_B T}{d_m  d_{\text{TC}}}$ is the surface energy per unit area \citep{raphael1992rubber, leger2008adhesion}, $k^c_+$ and $k^w_+$ are rate constants of cleavage and enzymatic unwinding, respectively, $N^s_{E_V E_R}$ is the number of enzymes at fully-bound state (one regular and one vulnerable sites), $\tau_c$ and $\tau_e$ are, respectively, a characteristic time and time required to form disentangled chains during detachment from the surface. See sections SI5 for  more details of  $F$, $\tau_c$ and $\tau_e$,  section SI6 for the reaction rate constants, and section SI7 for the initial conditions to solve the ODEs.  
We solved the system of ODEs using ODE23s in MATLAB (Mathworks, Natick, MA).

\subsection{Brownian dynamics (BD) to estimate the enzyme distribution around fibrils in a matrix}

\begin{figure}
\includegraphics[width=1\linewidth]{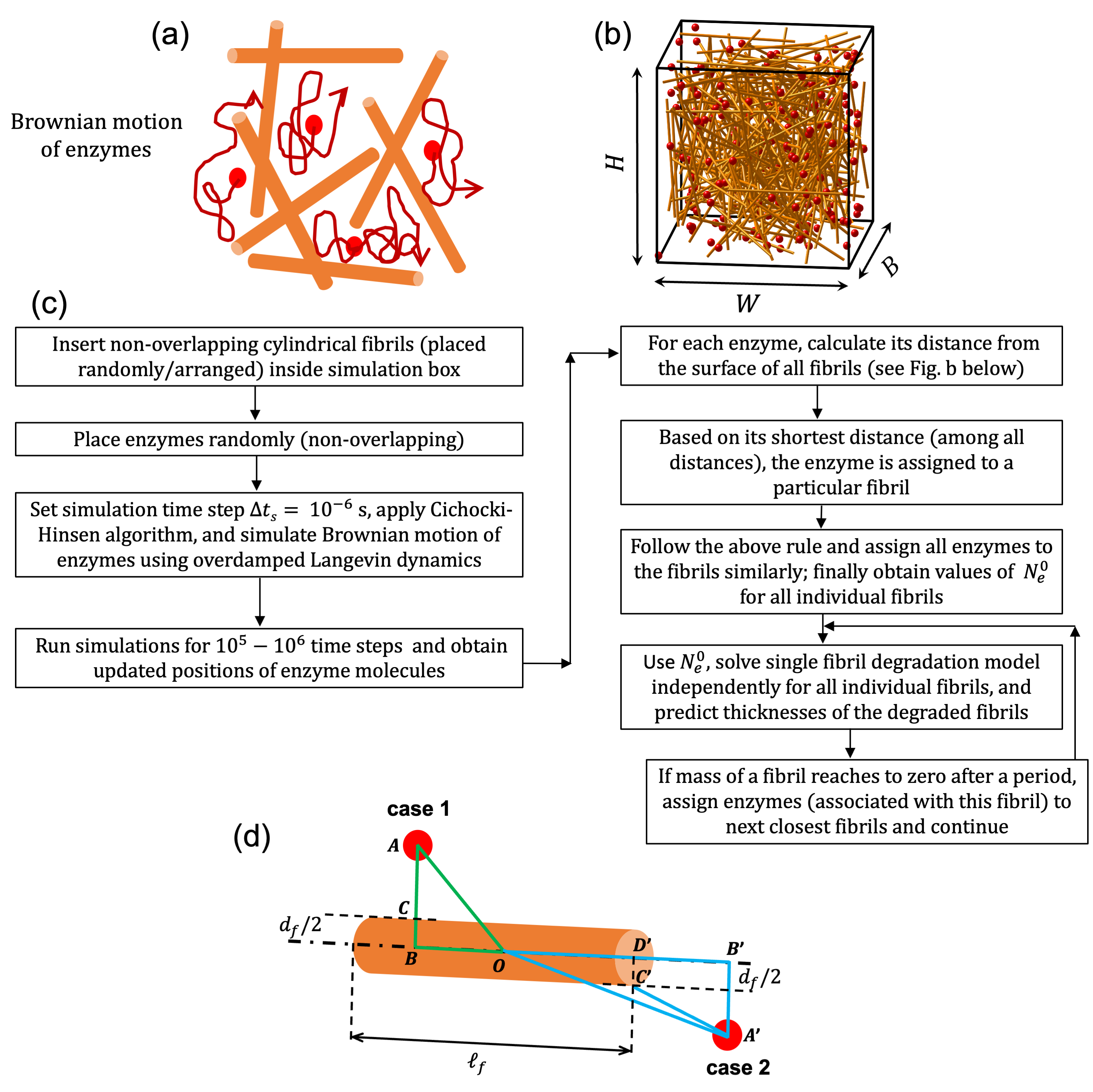}
\caption{{\textbf{Brownian dynamics and hybrid simulation.}} (a) Schematic of the Brownian motion of enzymes inside the pores.
(b) Configuration of the Brownian dynamics simulation where the cylinders represent collagen fibrils and  the red spheres represent the enzyme molecules (not to scale). (c) Flowchart of the hybrid modeling approach. (d) Schematic representing the distance of an enzyme from the surface of a fibril. Based on the position of the enzyme with respect to any fibril surface, two cases are possible. For case 1, the distance chosen is $A C = A B - d_f/2$. Otherwise, $A' C' = \sqrt{(B' D')^2 + (A' B' - d_f/2)^2}$  is the chosen distance for case 2, where $B' D' = \sqrt{(O A')^2 - (A' B')^2} - \ell_f/2$.}
 \label{fig:flowchart}
\end{figure}

We next set up a collagen matrix that consists of  stationary non-overlapping cylindrical fibrils (Fig.~\ref{fig:flowchart}a,b) for different  fibril fractions. 
We modeled the enzymes as spheres with a diameter of 10 nm \citep{tyn1990prediction}, and simulated their Brownian motion using overdamped Langevin dynamics \citep{ermak1978brownian}.
We used MATLAB to set up the simulations.
The fibrils and enzymes are inserted in the simulation box such that they do not overlap and are treated as hard particles.  The fibrils are randomly inserted using the methods given in  \cite{islam2016microstructure} and \cite{ayadfiber}.
Periodic boundary conditions are applied in all three directions. 
For the sake of simplicity, all potential interactions among fibrils and molecules are neglected and the  motion of the enzymes through the fibrous gel is simulated using Cichocki-Hinsen algorithm \citep{cichocki1990dynamic, smith2017fast, smith2017macromolecular}. 
The free diffusivity $D_e$ of the enzyme is set to $74 \times 10^{-12}$ m$^2$/s \citep{schultz2013monitoring}. The time step for simulations is chosen $\Delta t_s = 10^{-6}$ s such that it is  sufficiently small and the length increment in one time step is  $\sqrt{D_e \, \Delta t_s} \sim {\rm O}(d_E)$, which is equivalent to the size of the enzyme $d_E \sim 10$ nm. In simulations and model predictions, all length dimensions are scaled by $l_p = 1$ \textmu m, and the dimensions of the simulation box are chosen as $5 \, l_p \times 5 \, l_p \times 5 \, l_p$.
 The entire codebase used in this work along with a readme file is available at \url{https://github.com/RangamaniLabUCSD/Collagen_matrix_degradation.git}.

For a length scale $l_p = 1$ \textmu m and  the diffusivity of the  enzyme (collagenase) $D_e \sim 10^{-10}$ m$^2$/s \citep{schultz2013monitoring},  the time scale is  of O($l_p^2/D_e$) $\sim$ O($10^{-2}$) s.  Hence the limit of instantaneous diffusion is valid in the long time limit for  the chosen simulation box dimensions \citep{tzafriri2002reaction}. Based on this assumption, we proposed a hybrid framework (Fig.~\ref{fig:flowchart}c). In this framework, we run the simulation for a period of time and then obtain the number of enzymes ($N_e^0$) surrounding the fibrils based on the shortest distances of enzymes from the fibrils (Fig.~\ref{fig:flowchart}d). We  use the values of $N_e^0$ to solve the single fibril model where the fibril-scale model considers the binding-unbinding and other types of interactions among the enzymes and fibrils through a set of reactions.    Using this hybrid modeling framework, we predict the degradation of all individual fibrils, effectively the degradation of a matrix. 

To perform the BD simulations, we chose an enzyme concentration $\sim 2.5$ \textmu g/mL. For the simulation box volume $5 \, l_p \times 5 \, l_p \times 5 \, l_p = 125 \, l_p^3$, the weight of enzymes is $312.5 \times 10^{-18}$ g. The molecular weight ($M_w^e$) of collagenase varies between 70-130 kDa. We have chosen  $M_w^e \sim 120$ kDa \citep{eckhard2011structure}. For this chosen value, the estimate of total number of enzymes in simulation volume is $(N_e^0)_{total} \sim 1500$.

We fixed the enzyme concentration and vary the collagen concentration, or equivalently, the fibril fraction $\phi_f$. Here $\phi_f$ is the ratio of the volume of all fibrils to the volume of the simulation box. See section SI8 for more details on estimation of the fibril fraction $\phi_f$ from a collagen concentration. By varying  fibril diameter $d_f$,  fibril length $\ell_f$  and number of fibrils $n_f$, we vary $\phi_f$ in the range $0.003-0.03$, i.e. $0.3-3$\% for organs such as brain, liver, kidney, etc. \citep{tarnutzer2023collagen}. These three parameters, $d_f$, $\ell_f$ and $n_f$, are important factors to address different  matrix microarchitectures. The orientation and curvature (or, crimping) of the fibrils can be other factors \citep{ashworth2024importance}, however, we did not consider their roles in the present work.

\subsection{Experimental methods}

Following the experimental methodologies of \cite{ranamukhaarachchi2019macromolecular}, we used rat tail type I collagen to synthesize the collagen gels. To vary the microarchitecture of the gels, we used polyethylene glycol (PEG) as a macro-molecular crowding agent. After polymerization of collagen using PEG at 37$^{\text{o}}$C, PEG was washed out of the gels by rinsing them with Dulbecco's Modified Eagle Medium (DMEM) solution. We then treated the gels with a bacterial collagenase. To characterize the microarchitectures of the gels pre- and post-degradation, we performed fast green staining and imaging using the confocal fluorescence microscopy and scanning electron microscopy. See section SI9 for more details of the experimental methods.

\section{Results}
\subsection{Single fibril model captures experimentally observed degradation rates}

\begin{figure}
\centering
\includegraphics[width=0.6\linewidth]{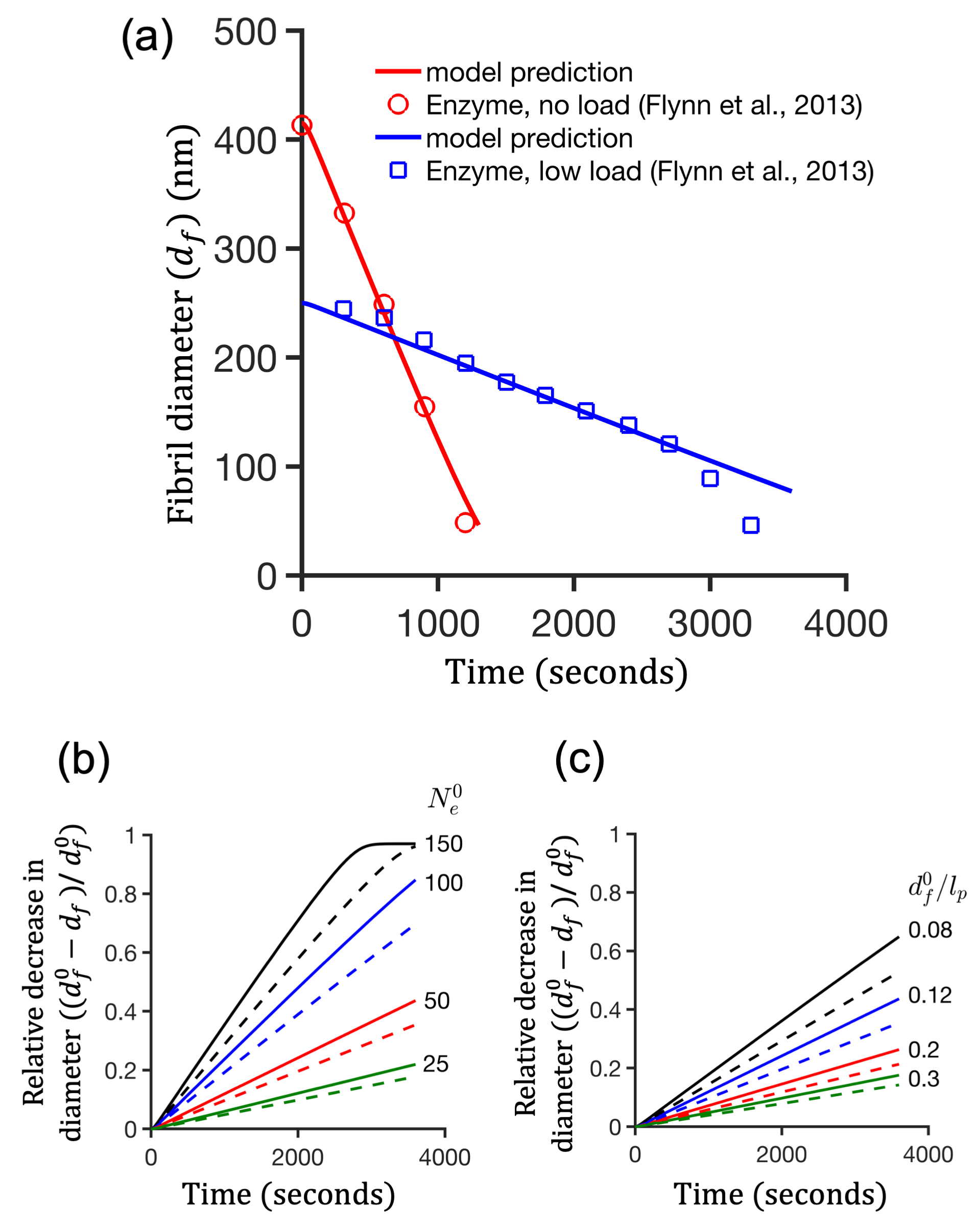}
\caption{{\textbf{Degradation of single fibril.}}   (a) Validation of the single fibril model  with  the experimental findings of \cite{flynn2013highly}.  Relative decrease of fibril diameter $(d_f^0-d_f)/d_f^0$ (or, the extent of degradation) with time for two cases: (b) fixed  initial fibril diameter $d_f^0/l_p = 0.12$   and varying  number of enzymes surrounding a fibril ($N_e^0$);  (c) Varying $d_f^0/l_p$  and fixed $N_e^0 = 50$. Here $l_p = 1$ \textmu m is a nominal length scale. The length of the fibril is $\ell_f/l_p = 2$ in (b) and (c). The solid and dashed curves in (b) and (c) correspond to $k^c_+ = $ 0.583 s$^{-1}$ \citep{mallya1992kinetics} and 0.472 s$^{-1}$ \citep{welgus1982gelatinolytic}, respectively. 
}
\label{fig:res_single_fibril}
\end{figure}

We first validated the reaction scheme for a single fibril surface erosion model against previously published experiments of \cite{flynn2013highly}. 
In the absence of  external loading, our model captured the experimental trend (Fig.~\ref{fig:res_single_fibril}a). See section SI10 for more  details related to the validation procedure. When the fibril is under external loading,  perhaps the external tension increases the stability of the triple helices by increasing the energy barrier for enzymatic unwinding \citep{chang2012molecular, chang2014molecular, tonge2015micromechanical, saini2020tension, topol2021fibrillar}, which is different from the  mechanism for an isolated triple helix under external tension \citep{adhikari2011mechanical,adhikari2012conformational}. Under low loading of $\sim$ 2 pN per tropocollagen monomer \citep{flynn2013highly,tonge2015micromechanical}, a small increase in the energy barrier $E_m$ for enzymatic unwinding by $\delta E_m = 0.013 \, E_m$ explains the experimental trend satisfactorily (Fig.~\ref{fig:res_single_fibril}a). See section SI10 for the choice of $\delta E_m = 0.013 \, E_m$. Overall, our model captured the experimental trends for a degrading fibril under different external loading conditions reasonably well.

In a surface erosion process, the  degradability of a fibril must be proportional to the ratio $N_e^0/A^0_f$, where $N_e^0$ and  $A^0_f = (\pi \, d_f^0 \, \ell_f)$ are the number of enzymes surrounding a fibril and the initial surface area of the fibril, respectively.  For a fibril of fixed initial diameter $d_f^0$, an increase in $N_e^0$ increases the degradability (Fig.~\ref{fig:res_single_fibril}b). However, for a fixed value of $N_e^0$, the degradability decreases with the increase in $d_f^0$   (Fig.~\ref{fig:res_single_fibril}c).  The quantity $(d_f^0 - d_f)/d_f^0$ represents the relative decrease in the diameter. Its smaller value represents less degradation and \textit{vice versa}. We provided a few results (Fig.~\ref{fig:res1_single_fibril}) on degradability and the ratio $N_e^0/A^0_f$ in section SI10. In summary, our model predictions captured the scaling related to a surface erosion process.

\subsection{Degradation of collagen matrices}

Having established that the single fibril model captures experimentally observed degradation rates (Fig.~\ref{fig:res_single_fibril}a), we next focused on degradation of collagen fibrils in matrices. Before addressing matrix degradation, we define two important dependent parameters: the initial   volume fraction of fibrils as $\phi_f = \Big( n_f \, \frac{\pi}{4} \, \sum_i ((d_f^0)^2 \, \ell_f)_i \Big)/ \text{vol.}^{\text {box}}$,  and the initial total surface area (scaled) of  fibrils as $(A_f^0)_{total} = \Big( n_f \, \pi \ \, \sum_i (d_f^0 \, \ell_f)_i \Big)/l_p^2$.

\subsubsection*{Degradation of matrices with  uniform fibrils}
\label{subsec:manyfibrils}

\begin{figure}
\centering
\includegraphics[width=1\linewidth]{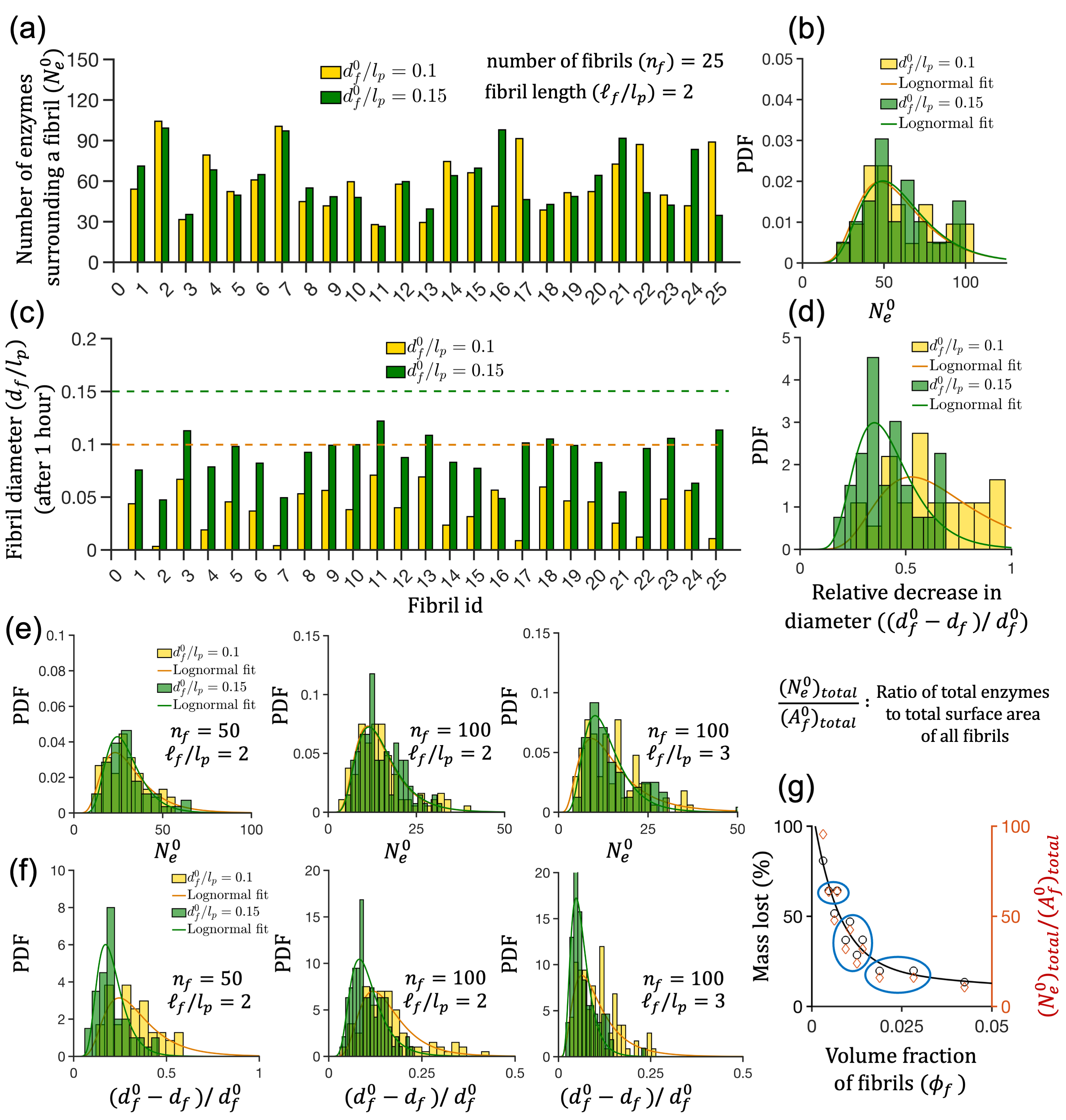}
\caption{{\textbf{Degradation of collagen matrices.}} 
(a)  The values of the number of enzymes $N_e^0$ surrounding the individual fibrils in a matrix for two different  matrices of same fibril length $\ell_f$ and number of fibrils $n_f$, but for two different values of initial fibril diameter $d_f^0/l_p = $ 0.1 and 0.15. 
(b) The probability density function (PDF) of $N_e^0$ generated using the data in (a).  (c) The predicted  diameters of the fibrils after 1 hour of degradation by enzymes for the cases in (a). (d) The  PDFs representing the relative decrease in diameter $(d_f^0 - d_f)/d_f^0$ for the data shown in (c).
Panels (e) and (f) represent PDFs of $N_e^0$ and $(d_f^0-d_f)/d_f^0$ for different values of $d_f^0$, $n_f$ and $\ell_f$. 
(g) The percentage of  the mass lost (left ordinate)  and the  ratio $(N_e^0)_{total}/(A_f^0)_{total}$ (right ordinate)  with the volume fraction of fibrils $\phi_f$. The black curve in (g) represents a fit to the data points corresponding to the left ordinate. The data points inside the blue circles show  non-monotonous trends of the mass lost and the ratio $(N_e^0)_{total}/(A_f^0)_{total}$ with respect to $\phi_f$. } 
\label{fig:res2_matrix}
\end{figure}

We first considered uniform fibrils of same initial diameter $d_f^0$ and length $\ell_f$, and investigated the effect of the number of fibrils $n_f$, $d^0_f$,  and $\ell_f$ on degradation.  Using the hybrid approach described in Fig.~\ref{fig:flowchart}b,c, we counted the number of enzymes $N_e^0$ surrounding every single fibril (Fig.~\ref{fig:res2_matrix}a).
We used this count of the number of enzymes to obtain a probability density function (PDF) of $N_e^0$  for a given matrix (Fig.~\ref{fig:res2_matrix}b).

\begin{figure}[hbtp]
\centering
\includegraphics[width=0.6\linewidth]{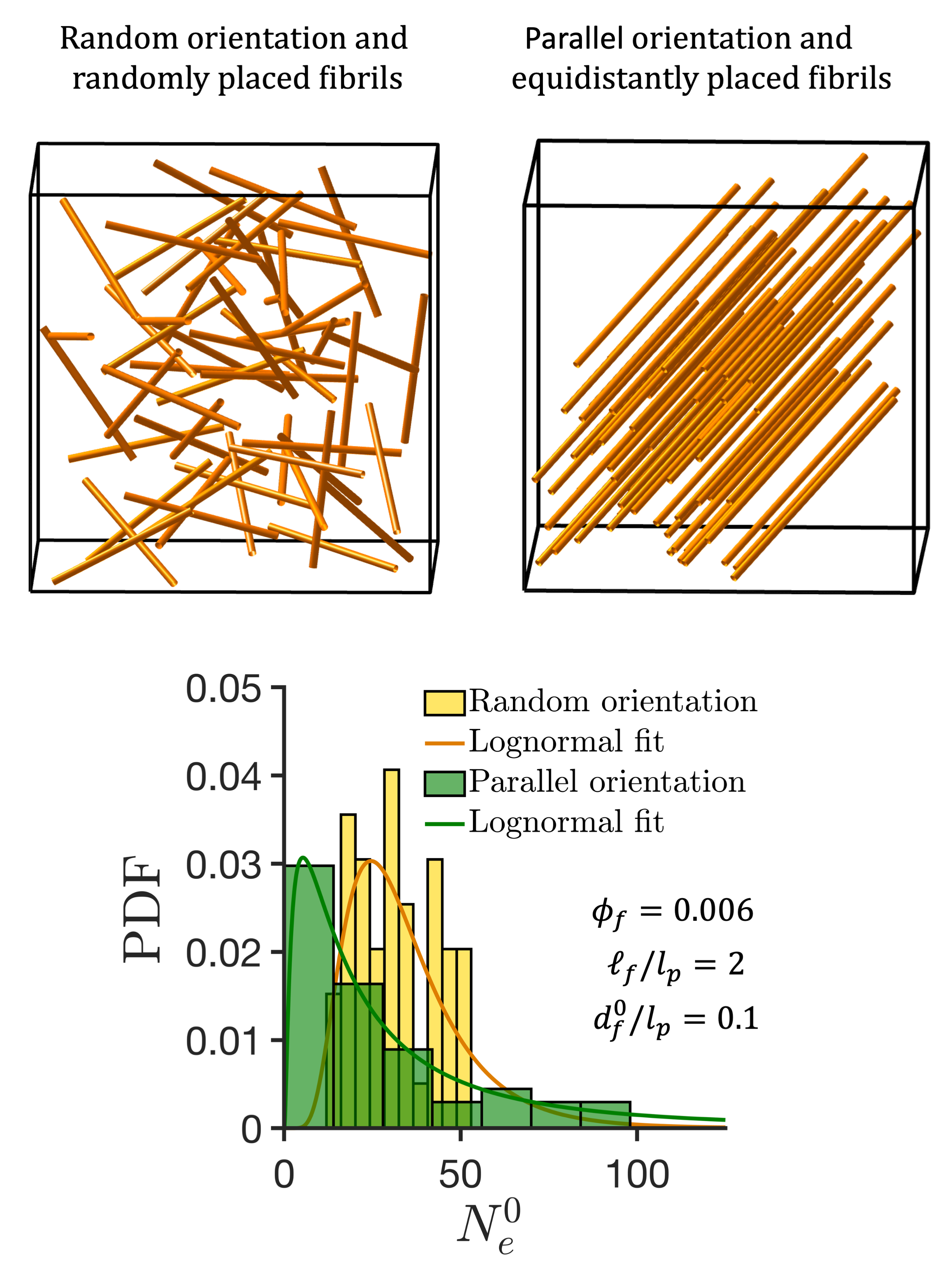}
\caption{The enzyme distributions in two different configurations.}
 \label{fig:en_dstr_config}
\end{figure}

We observed that the enzyme distribution was not uniform across the fibrils in a matrix (Fig.~\ref{fig:res2_matrix}a,b), suggesting that the organization of the collagen fibrils was a major determinant of the enzyme distribution. There are a few factors that can result in this non-uniform distribution of enzymes around fibrils. First, the cylindrical geometry of fibrils can result in an anisotropy of the diffusive motion of enzymes \citep{stylianopoulos2010diffusion,
chen2021noninvasive}.
Even if the cylindrical geometries of the fibrils are taken into account, for enzyme distributions in a matrix to be nearly uniform, fibrils should be organized in an equidistant manner with parallel alignment. 
This organization ensures a uniform pore size around cylindrical fibrils and can result in nearly uniform enzyme distribution (Fig.~\ref{fig:en_dstr_config}). 
However,  the random placement and random orientation of fibrils induces a non-uniform distribution of pore sizes. These factors together can result in a non-uniform enzyme distribution in a fibrillar matrix. Our simulations show that this finding of non-uniform distribution of enzymes holds for different values of fibril diameter, fibril length, and number of fibrils (Fig.~\ref{fig:res2_matrix}e).

The enzyme distributions do not differ much between two matrices if we vary only  $d_f^0$, fixing $\ell_f/l_p = 2$ and $n_f = 25$ (Fig.~\ref{fig:res2_matrix}b). Using the values of the number of enzymes in Fig.~\ref{fig:res2_matrix}a, we solved the single fibril model for each of the individual fibrils and calculated the diameters of the fibrils (Fig.~\ref{fig:res2_matrix}c).
As expected, each fibril degrades to a different extent because of the difference in the number of enzymes surrounding it. 
Using these single fibril data, we obtained the  PDF of the relative decrease in the  diameter of the fibril $(d_f^0-d_f)/d_f^0$ (Fig.~\ref{fig:res2_matrix}d). Thus, we find that although the initial fibril diameter is uniform, the degraded fibrils have a distribution of diameters because of the non-uniform enzyme distribution. 

We next investigated the effect of the number of fibrils $n_f$ on the extent of degradation.
From our simulations, we find that the enzyme distribution is a strong function of $n_f$ (Fig.~\ref{fig:res2_matrix}e). For a fixed value of $(N_e^0)_{total}$, we can expect a decrease in the average value of $N_e^0$ per fibril if $n_f$ increases.
When comparing two matrices of the same $d_f^0$ and  $\ell_f$, the matrix with higher number of fibrils degrades to a lesser extent than the matrix with fewer fibrils (Fig.~\ref{fig:res2_matrix}d,f). We also find that an increase in $d_f^0$ and $\ell_f$ decreases the degradability of the matrix for  fixed values of $n_f$ and $(N_e^0)_{total}$ (Fig.~\ref{fig:res2_matrix}d,f) because of the decrease in the ratio $(N_e^0)_{total}/(A_f^0)_{total}$.

\subsubsection*{Effect of matrix architecture on fibril degradation}

Since we model the single fibril degradation as a surface erosion process, the extent of degradation for a matrix must be proportional to the ratio of the total number of enzymes to the total surface area of the fibrils  $(N_e^0)_{total}/(A_f^0)_{total}$. 
The overall mass lost, defined as 
\begin{equation}
\text{mass lost} = \frac{\Big[\sum_{id=1}^{n{_f}} \, \frac{\pi}{4} \, (d_f^0)^2_{id}  \; \ell_f \Big] - \Big[\sum_{id=1}^{n{_f}} \, \frac{\pi}{4} \, (d_f)^2_{id}  \; \ell_f \Big]}{\sum_{id=1}^{n{_f}} \, \frac{\pi}{4} \, (d_f^0)^2_{id}  \; \ell_f} \times 100 \, ,
\label{eq:masslost}
\end{equation}
represents a direct estimate of the matrix degradation. For all combinations of $n_f$, $d_f^0$ and $\ell_f$, we find a direct correlation between  the mass lost and the ratio $(N_e^0)_{total}/(A_f^0)_{total}$  (Fig.~\ref{fig:res2_matrix}g) when they are plotted against the fibril fraction $\phi_f$.  However, the variation of  the mass lost and $(N_e^0)_{total}/(A_f^0)_{total}$ with $\phi_f$ are not strictly monotonous (blue circles in Fig.~\ref{fig:res2_matrix}g), although their trends appear to be monotonically decreasing as $\phi_f$ increases. 
The circled data points in Fig.~\ref{fig:res2_matrix}g show that for the same enzyme concentration, different extent of degradation can occur between two matrices of same fibril fraction $\phi_f$ (equivalently, the same collagen concentration) and {\textit{vice versa}}. 
This finding leads to the following question: does the difference in microarchitecture of two matrices of same $\phi$ impact the extent of degradation? 
We answer this question by conducting the following simulations. 

\begin{figure}
\centering
\includegraphics[width=1\linewidth]{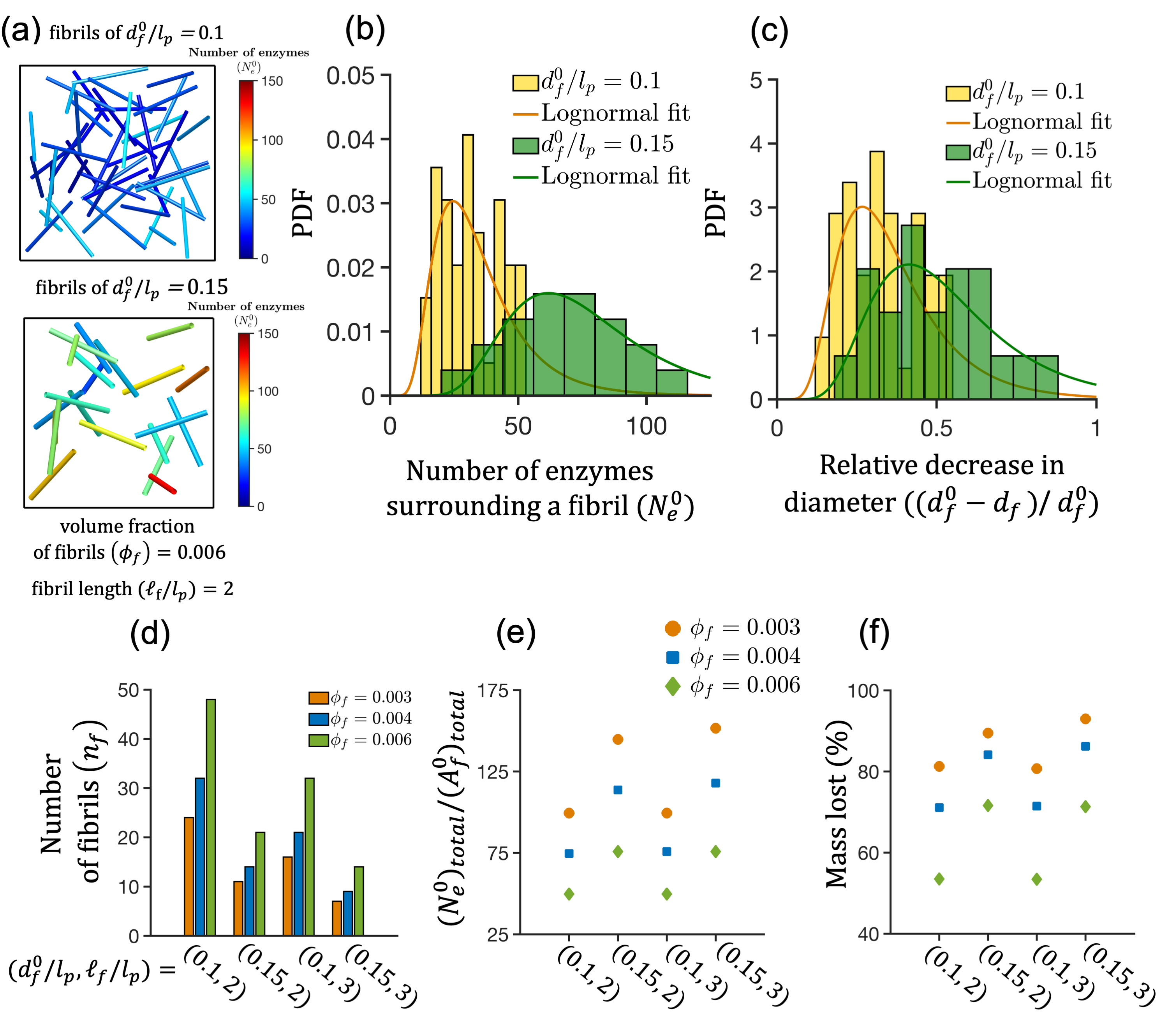}
\caption{{\textbf{Matrix with thicker fibrils degrades more than that with thinner fibrils.}} (a) The  initial configurations of two matrices having the same fibril fraction $\phi_f = 0.006$. The color bars in (a) represent the number of enzymes ($N_e^0$) surrounding the fibrils. For the configuration in (a),  the PDF of enzyme distribution $N_e^0$ (b) and   the PDF of the extent of degradation of fibrils $(d_f^0 - d_f)/d_f^0$ (c).  For matrices having different $\phi_f$,  the first (d), second (e) and third (f) panels report the number of fibrils in matrices, $(N_e^0)_{total}/(A_f^0)_{total}$ and the percentage mass lost (overall degradation of matrices), respectively.  All results correspond to the  total number of enzymes $(N_e^0)_{total} = 1500$ and  time 60 minutes.} 
\label{fig:res3_matrix}
\end{figure}

We fix the value of $\phi_f$ and the enzyme concentration, and consider the matrices with uniform fibrils. 
Without varying $\ell_f$, different matrices of same $\phi_f$ can be generated by adjusting the values of $n_f$ and $d_f^0$ (Fig.~\ref{fig:res3_matrix}a). 
For example, as $\phi_f \propto n_f \, (d_f^0)^2$, a 50\% decrease in $d_f^0$ can increase $n_f$ up to 4 times, resulting in twice the increase of the surface area ($(A_f^0)_{total} \propto n_f \, d_f^0$).  
For the case of constant $\phi_f$ and smaller $d_f^0$,  $n_f$ increases and shifts the enzyme distribution  towards the left (Fig.~\ref{fig:res3_matrix}b). 
Simulations show that a matrix composed of thinner fibrils degrades less than that a matrix composed of thicker fibrils (Fig.~\ref{fig:res3_matrix}c).
This is an unexpected result and counterintuitive to what we might expect from a single fibril model. 
In the single fibril degradation model, a thinner fibril degrades more than a thicker fibril if the enzyme concentration is the same (Fig.~\ref{fig:res_single_fibril}c)  because of lesser surface area of a thinner fibril available to larger number of enzymes.  In contrast, in a three-dimensional, randomly oriented and randomly placed fibril network, multiple factors affect the enzyme distribution as discussed previously. 
Thus, this finding highlights the importance of incorporating the three-dimensional spatial considerations including diffusion of enzymes and matrix microarchitectures into the model. 
The results in  Fig.~\ref{fig:res3_matrix}d-f show how the number of fibrils changes for different sets of ($d_f^0$, $\ell_f$) when $\phi_f$ is fixed, which directly influences the ratio $(N_e^0)_{total}/(A_f^0)_{total}$, and  the overall mass lost.
In summary, our simulations predict that for the same enzyme concentration, uniform fibril diameter, and the same fibril fraction $\phi_f$, a matrix with thicker fibrils can degrade more than that with thinner fibrils  (Figs.~\ref{fig:res3_matrix}d-f).

To compare this model prediction on matrix degradability, we performed {\it in vitro} experiments with synthetic collagen gels and investigated their degradability.

\subsection{Experiments reveal that matrix microarchitecture governs degradability}

\begin{figure}
\centering
\includegraphics[width=1\linewidth]{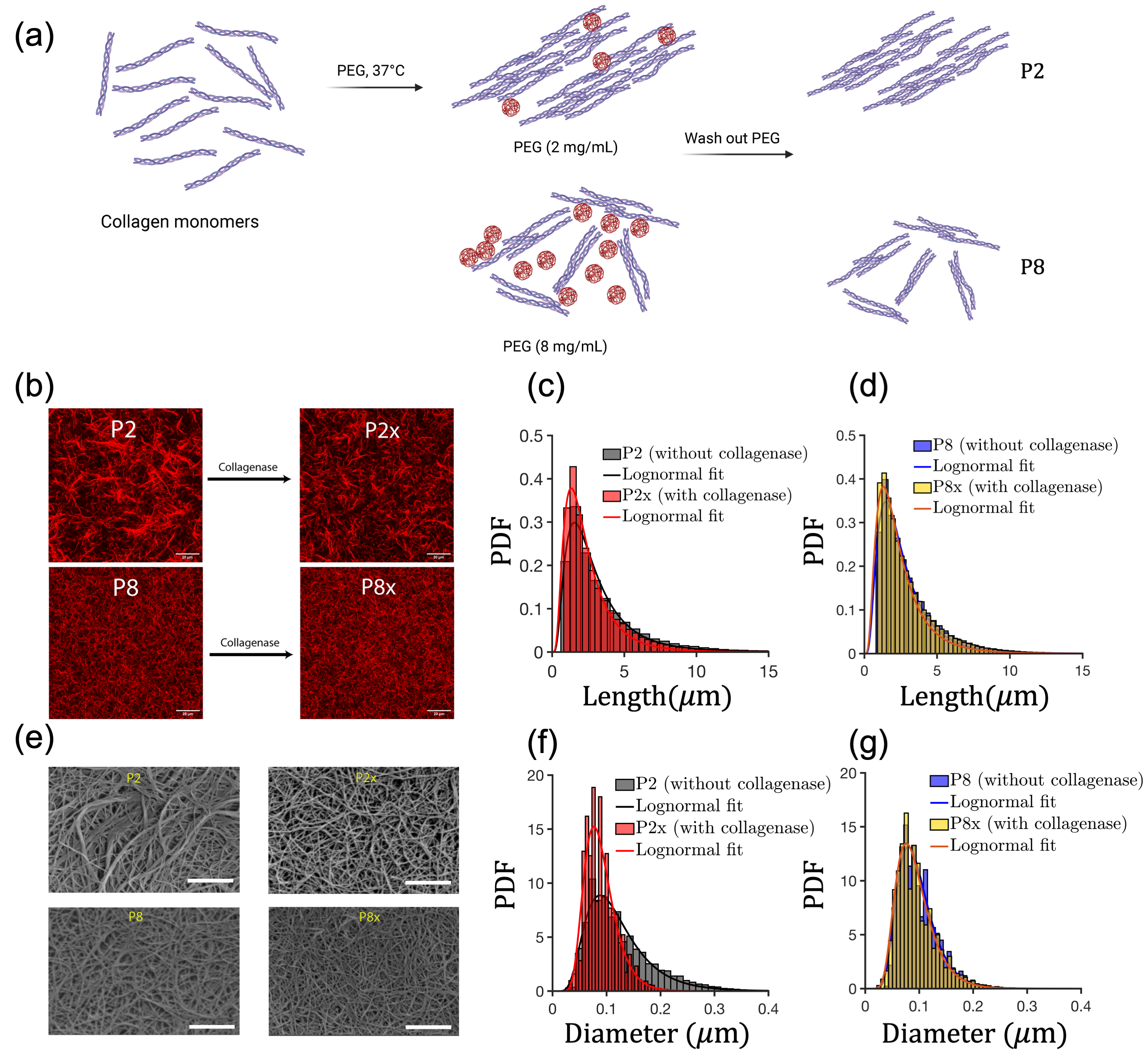}
\caption{{\textbf{Experimental findings.}} (a) Preparation of collagen gels. (b) The fast green stained images of the gels (scale bar 20 $\mu$m). The length distributions of (c) P2/P2x and (d) P8/P8x before and after treatment with collagenase from the green stained images. (e) The SEM images of P2/P2x and P8/P8x (scale bar 2 $\mu$m). The histograms of fibril diameters for (f) P2/P2x and (g) P8/P8x.} 
\label{fig:res_exp}
\end{figure}

\cite{ranamukhaarachchi2019macromolecular} previously showed that the microarchitecture of a collagen gel can be modulated using polyethylene glycol (PEG) as a macro-molecular crowding (MMC) agent. 
They also reported that the degradability of the gels varies with the matrix microarchitecture. 
Using this background, to test our model predictions, we prepared collagen gels with two different microarchitectures from the same collagen concentration. We used 2.5 mg/ml collagen (final concentration) and two different MMC concentrations: 2 mg/mL PEG designated as P2 and 8 mg/mL PEG denoted as P8 (Fig.~\ref{fig:res_exp}a). We used a bacterial collagenase concentration of  2.5 \textmu g/mL  (final concentration) to perform degradation experiments. 
Using fast green staining images and scanning electron microscopy (SEM), we quantified the fibril length and thickness distributions respectively before and after degradation. 
We denote the gels post-degradation as P2x and P8x.  

Fig.~\ref{fig:res_exp}b shows the fast green stained images, and the length distributions for P2/P2x (Fig.~\ref{fig:res_exp}c) and P8/P8x (Fig.~\ref{fig:res_exp}d). The mean length of the fibrils in P8 ($\sim 3 \pm 0.2$ \textmu m) is slightly smaller than that of P2 ($\sim 3.3 \pm 0.4$ \textmu m).
From Fig.~\ref{fig:res_exp}b,  it is obvious that P2 degrades more than P8 as larger pores are present in P2x. However, there is no significant change in the length distributions before and after degradation in both P2/P2x and P8/P8x (Fig.~\ref{fig:res_exp}c,d); the decrease in the mean length post-degradation is less than $\sim 8$\%. 
Hence the assumption of treating the fibril length $\ell_f$ as a constant in our model is supported by the  experimental findings.
The green stained images  (Fig.~\ref{fig:res_exp}b) reveal that the number of fibrils is higher in P8 than that in P2 (through visualization).

The SEM images (Fig.~\ref{fig:res_exp}e) and the histograms of the fibril diameters (Fig.~\ref{fig:res_exp}f,g) show that the fibrils are thicker in P2 than in P8. The mean diameter of the fibrils in P8 ($\sim 0.09 \pm 0.02$ \textmu m) is  $\sim$ 25-30\% smaller than that of P2 ($\sim 0.13 \pm 0.03$ \textmu m).  After degradation, the decrease in the diameter occurs in both P2x and P8x (Fig.~\ref{fig:res_exp}f,g).  Quantification of the diameters of the fibrils show that P2 (Fig.~\ref{fig:res_exp}f) degrades more than P8 (Fig.~\ref{fig:res_exp}g).
The decrease in the mean diameter is $\sim 30-40$\% in P2x and $\sim$ 15\% in P8x.   
Thus, experiments validate our model predictions that a matrix with thicker fibrils can degrade more than one with thinner fibrils for the same collagen and collagenase concentrations.

\subsubsection*{Model predictions for matrices with non-uniform fibrils highlight the role of microarchitecture in degradation}

To further reinforce the role of matrix microarchitecture, we note that the significant differences between  the microarchitectures of P2 and P8 are primarily due to the number of fibrils and fibril diameter. 
In P2, there are less number of fibrils and the fibrils are thicker. However, in P8, there are higher number of fibrils  but the fibrils are thinner, compared to P2 (Fig.~\ref{fig:res_exp}). 
Our model predicts higher degradability of a matrix with thicker fibrils than that with thinner fibrils (Figs.~\ref{fig:res3_matrix}). Thus, we can qualitatively explain why the matrix with the P2 architecture degrades more than the P8 architecture. However, our simulations  in Fig.~\ref{fig:res3_matrix} are for fibrils of uniform initial diameter. 
Because the experimentally synthesized matrices had fibrils with a distribution of diameters, we next simulated the degradation of fibrils in  matrices of experimentally-inspired fibril diameter distributions. 

\begin{figure}[hbtp]
\centering
\includegraphics[width=1\linewidth]{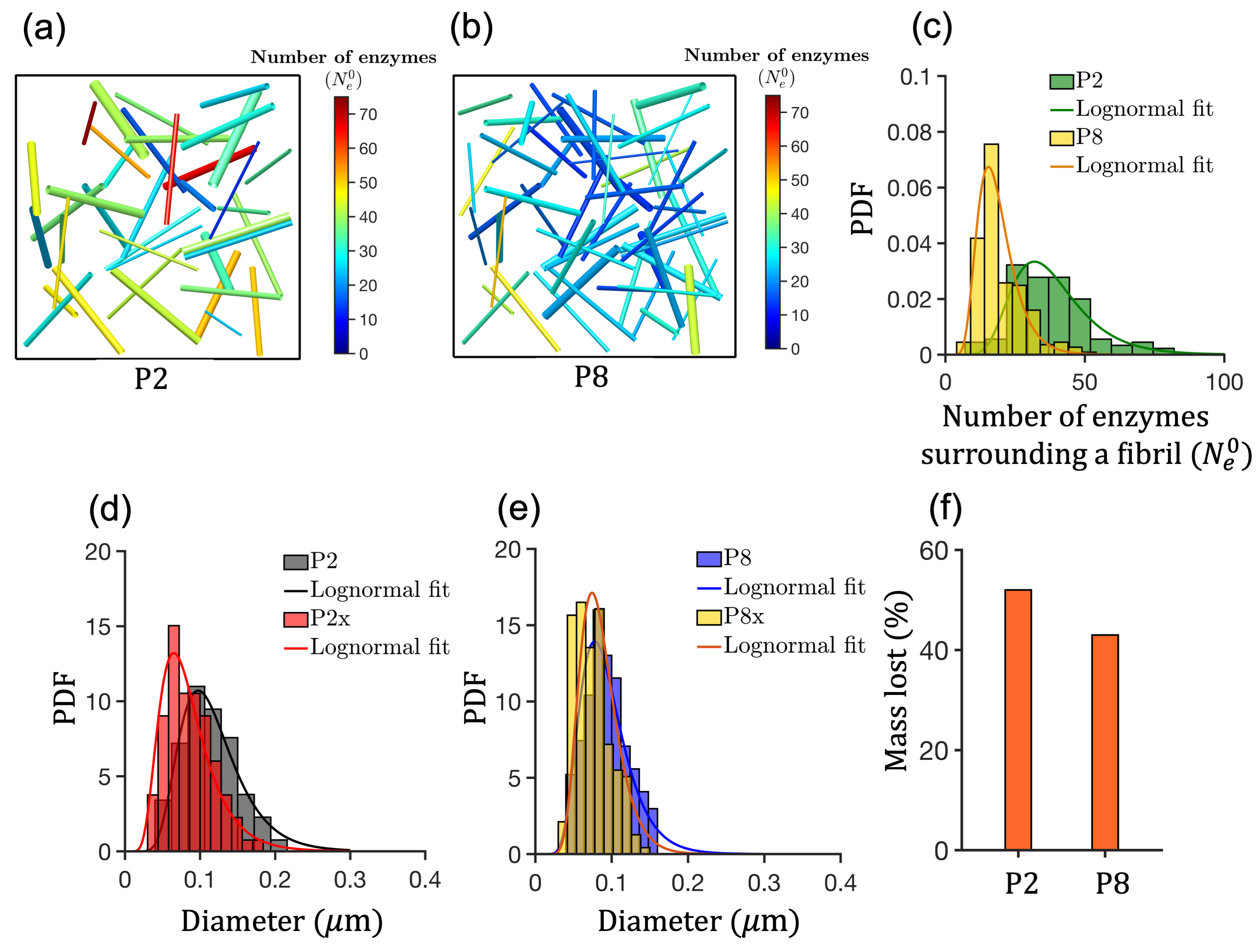}
\caption{{\textbf{Simulations of matrices with non-uniform fibrils predict that P2 degrades higher than P8.}}  The  initial configurations of two matrices, P2 (a) and P8 (b), with the same fibril fraction $\phi_f = 0.007$. The color bars in (a) and (b) represent the number of enzymes ($N_e^0$) surrounding the fibrils. For the configuration in (a) and (b),  the PDF of the enzyme distributions $N_e^0$ (c),   the PDF of the diameters of the fibrils for P2/P2x (d) and P8/P8x (e), and the overall mass lost (f).  The results correspond to the total number of enzymes $(N_e^0)_{total} = 1500$ and  time 60 minutes.} 
\label{fig:res_sim_compare}
\end{figure}

We generated the fibrils whose diameters matched the experimentally observed diameters of P2 and P8 (Fig.~\ref{fig:goodness_fits} and Fig.~\ref{fig:matrix_gen}) while maintaining the fibril length to be the same (see section SI11 for more details). 
We used a fibril fraction of $\phi_f = 0.007$ in both cases as this value is close to the collagen concentration used in the experiments. As a result, in these newly generated matrices, the number of fibrils  is higher for P8 ($n_f \sim 75$) than that in P2 ($n_f \sim 40$) for the same fibril fraction $\phi_f$. We compared the outcomes of degradation in these conditions as shown in Fig.~\ref{fig:res_sim_compare}. The microarchitectures are  different for P2 and P8 (Fig.~\ref{fig:res_sim_compare}a,b) in terms of the number of fibrils and the diameters. 
As a result, the enzyme distribution for P8 shifts to the left due to larger number of fibrils (Fig.~\ref{fig:res_sim_compare}c) and implies less number of enzymes per fibril.  The diameter distributions (in  the range 0.03-0.2 \textmu m) of P2 and P8 before and after degradation  (Fig.~\ref{fig:res_sim_compare}d,e) and the overall mass lost (Fig.~\ref{fig:res_sim_compare}f) indicate that P2 degrades more than P8 for the same collagen and enzyme concentrations, in agreement with experiments (Fig.~\ref{fig:res_exp}e,f). Our simulation results using non-uniform fibril diameter only reinforce our model predictions that the matrices of the same collagen concentration can have very different microarchitectures and the matrices can degrade differently under the same collagenase concentration.  In summary, our study reveals that matrix microarchitecture plays an important role in degradation of collagen matrices.

\section{Discussion}
\label{sec:discussion}

In this study, using a combination of modeling and experiments, we investigated the collagenolytic degradation of collagen matrices. We showed that the matrix microarchitecture is a strong determinant of matrix degradability.  The enzyme distribution in a matrix is not uniform due to the fibrillar network, and this enzyme distribution leads to fibrils of distributed diameters even if  the fibrils are initially uniform in size. The matrices of the same collagen concentration can degrade differently under the same collagenase concentration because of the differences in the  microarchitecture. These findings are backed by {\it in vitro}  experiments with collagen gels of different microarchitectures. 

Unraveling the connection between the matrix microarchitecture and degradation is important to design biomaterials and understand ECM remodeling at cellular length scale. The previous models are either continuum or discrete, and thus, they predict either overall rate of degradation or degradation at nanoscale \citep{tzafriri2002reaction,metzmacher2007model,vuong2017biochemo,
sarkar2012single}.  The model and framework we developed in this work to investigate the collagen matrix microarchitecture and matrix degradation  is novel in the sense that the previous models are not capable to address the connection between the degradation and microarchitectures. However, our model  has  also some drawbacks. The lattice-based single fibril model  has a limitation to predict the spatial variations in the size and shape of a degrading fibril. In our simulations, the fibrils are non-stretchable and  we did not consider explicitly the potential energy based attractive interactions among the fibrils and enzymes.  
At present, it is difficult to state {\textit{a priori}} whether  the potential interactions can affect the temporal evolution of  the enzyme distribution which merits further attention.

In summary, our single fibril model and hybrid modeling framework effectively capture multi-scale effects to  predict the degradation of three-dimensional matrices with different microarchitectures. Although relatively simple, this framework sheds light on how collagen matrix degradability is tuned by matrix microarchitecture. This has important implications for a number of fields, including matrix biology and biomaterials.

\section*{Conflicts of interest}
There are no conflicts to declare.

\section*{Acknowledgements}
 This work was supported by a National Science Foundation grant DMS-1953469, American Cancer Society Research Scholar Grant RSG-21-033-01-CSM to S.I.F., the National Cancer Institute U54CA274502, a Prebys Research Heroes Grant to S.I.F.  P.R. and S.I.F were also supported by the Wu Tsai Human Performance Alliance and the Joe and Clara Tsai Foundation. We would like to thank the UC San Diego School of Medicine Microscopy Core, which is supported by the National Institute of Neurological Disorders and Stroke grant P30NS047101.

\newpage
\bibliographystyle{apa}
\bibliography{References}

\newpage

\section*{Supplementary Information (SI)}
\beginsupplement

\subsection{Reaction scheme:}

\newpage
\subsection{Adsorption kinetics and $\Phi_{ad}$:}

\noindent
For reversible partial binding (represented by  eqn (\ref{eq:kinetics1})), we implemented the protein adsorption-desorption kinetics \citep{adamczyk1999irreversible, fang2005kinetics}. The rate of adsorption is proportional to the available surface function $\Phi_{ad} = 1 - (\theta/\theta_{max})$. This available surface function is related to the steric hindrance during  adsorption \citep{adamczyk1999irreversible, adamczyk2000kinetics, fang2005kinetics}, where $\theta$ and $\theta_{max}$ are the surface coverage fraction and its maximum saturated value, respectively. If the number of total enzymes is very small compared to the number of available sites, $\Phi_{ad}(\theta) \rightarrow 1$ in the limit of very low surface coverage. If the limit of low surface coverage does not hold true, some descriptions related to $\Phi_{ad}(\theta)$ are provided in earlier work \citep{schaaf1989surface, talbot2000car, adamczyk2000kinetics}. In our work, the limit $N^s \gg N_{EE}(t = 0)$ holds true for  low concentration of enzyme, where $N^s$ and $N_{EE}$ are total number of sites exposed at the surface and number of enzymes, respectively. Thus we  set $\Phi_{ad} = 1$.

\subsection{Correction factor $\phi$ in the intrinsic rates of full-length binding kinetics:} 

\noindent
In our model, we treated the forward rates of eqn (\ref{eq:kinetics2})-(\ref{eq:kinetics3})  as pseudo first-order kinetics multiplied with a correction factor $\phi$. The functional form of $\phi$  increases with the number ($n_a$) of available lattice sites per enzyme at partially bound state. The value of $\phi$ is zero if there is no  site available, $\phi = 1$  when only one lattice site is available ($n_a = 1$), and $\phi$ must saturate around $(\pi \, (d_E)^2)/(d_m  \, d_{\text{TC}})$ which is equivalent to the number of lattice sites inside the searching radius $d_E$ (see Fig.~\ref{fig:schematic_model}c). We proposed a phenomenological function for $\phi$ as
\begin{equation}
    \phi  \approx \frac{\pi \, (d_E)^2}{d_m \, d_{\text{TC}}} \; \frac{n_a}{c_1 + n_a},
    \label{eq:correction_phi_supp}
\end{equation}
where $n_a = N^s/(N^s_{E_R E} + N^s_{E_V E})$, $N^s$ is the number of available sites on the fibril surface, $N^s_{E_R E}$ and $N^s_{E_V E}$ are the number of enzymes partially bound to one regular site and one vulnerable site, respectively. Here $c_1 = 25$ is a dimensionless constant which is obtained by setting $\phi(n_a = 1) = 1$ for $d_E = 10$ nm,  $d_{\text{TC}} = 1.5$ nm and $d_m = 8$ nm.

\subsection{System of ODEs:} 

\noindent
The system of ODEs representing the reaction scheme eqn (\ref{eq:kinetics1})-(\ref{eq:kinetics5}) is the following
\begin{align}
\frac{\md N_{EE}}{\md t} \, =  & \, - \, k^1_+ \, N_{EE} \, \Phi_{ad}(\theta) \,  + \, k^1_- \, N^s_{E_R E} \, - \, k^2_+ \, N_{EE} \, \Phi_{ad}(\theta) \,  + \, k^2_- \, N^s_{E_V E} \,    \notag\\
& + \, k^c_+ \, N^s_{(E_V E_R)^*}  \label{eq:enzyme}\\ 
\frac{\md N^s_R}{\md t} \, =  & \, - \, k^1_+ \, N_{EE} \, \Phi_{ad}(\theta) \,  + \, k^1_- \, N^s_{E_R E} \,  -  \, k^3_+ \, N^s_{E_R E} \, \phi \,  + \, k^3_- \, N^s_{E_R E_R}  \,  \notag \\
& - \, k^4_+ \, N^s_{E_V E} \, \phi \, - \, \Re   \label{eq:regular} \\
\frac{\md N^s_V}{\md t} \,  =  & \, - \, k^2_+ \, N_{EE} \, \Phi_{ad}(\theta) \,  + \, k^2_- \, N^s_{E_V E} \, - \, k^5_+ \, N^s_{E_R E}   \label{eq:vulnerable} \\
\frac{\md N^s_{E_R E}}{\md t} \,  =  & \,  \, k^1_+ \, N_{EE} \, \Phi_{ad}(\theta) \,  - \, k^1_- \, N^s_{E_R E} \, - \, k^3_+ \, N^s_{E_R E} \, \phi \, + \, k^3_- \, N^s_{E_R E_R} \,   \notag \\ 
& - \, k^5_+ \, N^s_{E_R E}    \label{eq:partial_regular_bind}  \\
\frac{\md N^s_{E_V E}}{\md t} \,  =  & \,  \, k^2_+ \, N_{EE} \, \Phi_{ad}(\theta) \,  - \, k^2_- \, N^s_{E_V E} \, - \, k^4_+ \, N^s_{E_V E} \, \phi  \label{eq:partial_vulnerable_bind} \\
\frac{\md N^s_{E_R E_R}}{\md t} \,  =  & \,  \, k^3_+ \, N^s_{E_R E} \, \phi \,  - \, k^3_- \, N^s_{E_R E_R}   \label{eq:full_regular_bind} \\
\frac{\md N^s_{E_V E_R}}{\md t} \,  =  &   \, k^4_+ \, N^s_{E_V E} \, \phi \,  + \, k^5_+ \, N^s_{E_R E}   \, - \, k^w_+ \, N^s_{E_V E_R}   \label{eq:full_vulnerable_bind}  \\ 
\frac{\md N^s_{(E_V E_R)^*}}{\md t} \,  =  &  \, k^w_+ \, N^s_{E_V E_R}  \, - \, k^c_+ \, N^s_{(E_V E_R)^*}  \label{eq:unwinding} \\
\frac{\md N_P}{\md t} \,   =   &  \, k^c_+ \, N^s_{(E_V E_R)^*} \,  + \, \frac{1}{2}  \Re  \, ,    \label{eq:product}  
\end{align}
where $N_{EE}$, $N^s_R$, $N^s_V$, $N^s_{E_R E}$, $N^s_{E_V E}$, $N^s_{E_R E_R}$, $N^s_{E_V E_R}$, $N^s_{(E_V E_R)^*}$, $N_P$ are numbers of free enzymes, regular sites, vulnerable sites, enzymes partially bound to one regular site, enzymes partially bound to one vulnerable site, enzymes fully bound to two regular sites, enzymes fully bound to one regular and one vulnerable site, enzymes fully bound to one regular and one vulnerable site (unwound state), and product sites, respectively. Here $k^1_+$, $k^1_-$, $k^2_+$, $k^2_-$, $k^3_+$, $k^3_-$, $k^4_+$, $k^5_+$, $k^w_+$, and $k^c_+$ are the reaction rate constants having dimension [time]$^{-1}$; $\Phi_{ad}$ and $\phi$ are available surface function and correction factor related to adsorption kinetics and full-length binding/hopping kinetics, respectively.   We  included another  term $\Re$  in the series of ODEs (eqn (\ref{eq:enzyme})-(\ref{eq:product})) in an \textit{ad hoc} manner, which is related to force-assisted removal of regular sites. We  discussed about $\Re$  in the main text and in the next section. The balances for the  total number of enzymes and total number of sites are the following.  

\vspace{0.3cm}
\noindent
{\textbf {Balance for total number of enzymes:}}
\begin{align}
\notag
&\frac{\md N_{EE}}{\md t} + \frac{\md N^s_{E_R E}}{\md t} + \frac{\md N^s_{E_V E}}{\md t} + \frac{\md N^s_{E_R E_R}}{\md t} + \frac{\md N^s_{E_V E_R}}{\md t} + \frac{\md N^s_{(E_V E_R)^*}}{\md t}  \\  \notag
 & = \Big[ \, - \, k^1_+ \, N_{EE} \, \Phi_{ad}(\theta) \,  + \, k^1_- \, N^s_{E_R E} \, - \, k^2_+ \, N_{EE} \, \Phi_{ad}(\theta) \,  + \, k^2_- \, N^s_{E_V E} \, + \, k^c_+ \, N^s_{(E_V E_R)^*} \,  \Big] \\  \notag
 & + \Big[  \, k^1_+ \, N_{EE} \, \Phi_{ad}(\theta) \,  - \, k^1_- \, N^s_{E_R E} \, - \, k^3_+ \, N^s_{E_R E} \, \phi \, + \, k^3_- \, N^s_{E_R E_R} \, - \, k^5_+ \, N^s_{E_R E} \Big] 
 \\   \notag
 & + \Big[  \, k^2_+ \, N_{EE} \, \Phi_{ad}(\theta) \,  - \, k^2_- \, N^s_{E_V E} \, - \, k^4_+ \, N^s_{E_V E} \, \phi \Big] \, + \, 
\Big[  \, k^3_+ \, N^s_{E_R E} \, \phi \,  - \, k^3_- \, N^s_{E_R E_R} \, \Big]  \\   \notag
& + \, \Big[  \, k^4_+ \, N^s_{E_V E} \, \phi \,  + \, k^5_+ \, N^s_{E_R E}   \, - \, k^w_+ \, N^s_{E_V E_R} \Big] \, + \, \Big[  k^w_+ \, N^s_{E_V E_R}  \, - \, k^c_+ \, N^s_{(E_V E_R)^*} \Big]\\  
& = \, 0
\label{eq:bal_enzymes}
\end{align}

\vspace{0.3cm}
\noindent
{\textbf {Balance for total number of lattice sites:}}
\begin{align}
\notag
&\frac{\md N^s_R}{\md t} + \frac{\md N^s_V}{\md t} + \frac{\md N^s_{E_R E}}{\md t} + \frac{\md N^s_{E_V E}}{\md t} + 2 \, \frac{\md N^s_{E_R E_R}}{\md t} + 2\, \frac{\md N^s_{E_V E_R}}{\md t} + 2  \frac{\md N^s_{(E_V E_R)^*}}{\md t}  + 2  \frac{\md N_p}{\md t}
\\  \notag
& = \Big[ - \, k^1_+ \, N_{EE} \, \Phi_{ad}(\theta) \,  + \, k^1_- \, N^s_{E_R E} \,  -  \, k^3_+ \, N^s_{E_R E} \, \phi \,  + \, k^3_- \, N^s_{E_R E_R}  \, 
 - \, k^4_+ \, N^s_{E_V E} \, \phi \,   -  \, \Re   \Big] \\  \notag
 & + \, \Big[ - \, k^2_+ \, N_{EE} \, \Phi_{ad}(\theta) \,  + \, k^2_- \, N^s_{E_V E} \, - \, k^5_+ \, N^s_{E_R E}  \Big] \\  \notag
 & + \, \Big[ k^1_+ \, N_{EE} \, \Phi_{ad}(\theta) \,  - \, k^1_- \, N^s_{E_R E} \, - \, k^3_+ \, N^s_{E_R E} \, \phi \, + \, k^3_- \, N^s_{E_R E_R} \, - \, k^5_+ \, N^s_{E_R E}  \Big] \\ \notag
 & + \, \Big[k^2_+ \, N_{EE} \, \Phi_{ad}(\theta) \,  - \, k^2_- \, N^s_{E_V E} \, - \, k^4_+ \, N^s_{E_V E} \, \phi \Big] \\ \notag
 & + \, 2 \Big[k^3_+ \, N^s_{E_R E} \, \phi \,  - \, k^3_- \, N^s_{E_R E_R} \Big] \, + \, 2 \Big[ k^4_+ \, N^s_{E_V E} \, \phi \,  + \, k^5_+ \, N^s_{E_R E}  \, - \, k^w_+ \, N^s_{E_V E_R} \Big] \\ \notag
 & + \, 2 \Big[ k^w_+ \, N^s_{E_V E_R}  \, - \, k^c_+ \, N^s_{(E_V E_R)^*} \Big] \, + \, 2 \Big[ k^c_+ N^s_{(E_V E_R)^*}  \,  + \, \frac{1}{2}  \Re \Big] \\
& = \, 0
\label{eq:bal_sites}
\end{align}


\newpage
\subsection{Force-assisted removal and description related to $\Re$:} 

\noindent
We assumed that the energy barrier for temporary winding-unwinding of a lattice site due to thermal fluctuations is symmetric in absence of enzymes.   Both the rates to cross the energy barrier from either side  are proportional to ${\rm exp}\,(-E_m/(k_B T))$ resulting in zero net rate, where $E_m$ is the energy required to cross the barrier, $k_B$ is the Boltzmann constant, and $T$ is the temperature. Due to a force $F$ emerging from enzyme-induced unwinding \citep{eckhard2011structure}, the energy barrier for other regular sites can become asymmetric. The energy required for the transition towards unwinding is reduced to $(E_m - \lambda_m F)$, and for the other side, it is increased to $(E_m + \lambda_m F)$. Here $\lambda_m$ is the extent of dissociation in unwound state which is chosen as  $ \sim 3.6 $ \AA  
   $\,$ \citep{perumal2008collagen}. Thus the net rate of flow over the energy barrier towards unwinding according to the theory of reaction rates is \citep{glasstone1941theory, stuart1953dependence}
\begin{equation}
    k_R = \, \,  \frac{k_B T}{h} \, \Bigg[{\rm exp} \Bigg(\frac{-(E_m - \lambda_m F)}{k_B T} \Bigg) - {\rm exp} \Bigg(\frac{-(E_m + \lambda_m F)}{k_B T} \Bigg) \Bigg],
    \label{eq:rate_theory1}
\end{equation}
where $h$ is Planck constant. For $F = 0$, the form  (\ref{eq:rate_theory1}) yields to zero. Here the value of $E_m$ is chosen as $E_m = n_{res} \, \Delta G/N_A$, where $\Delta G = 1.9$ kJ/mole per amino acid residue, $N_A$ is Avogadro number, and $n_{res} = 28$ is the number of amino acid residues corresponding to one lattice site \citep{perumal2008collagen}.  Thus the rate of force assisted removal of exposed regular sites is 
\begin{equation}
 \Re = k_R \, N^s_R ,
 \label{eq:rate_theory2}
\end{equation}
which is added in an \textit{ad hoc} manner in the system of ODEs (eqn (\ref{eq:enzyme})-(\ref{eq:product})). We derive the force $F$ in a heuristic manner. 

The force corresponding to the local stress generated during enzyme-induced unwinding at one lattice (vulnerable) site is equivalent to the force necessary to generate a new surface via slippage of the chain on the surface \citep{adjari1994slippage, sung1995slippage,  vega2001terminal}, i.e.
\begin{equation}
    f \sim \gamma \, v_s \, \tau_c ,
    \label{eq:force_der1}
\end{equation}
where $\gamma \sim \frac{k_B T}{d_m  d_{\text{TC}}}$ is the surface energy per unit area \citep{raphael1992rubber, leger2008adhesion}, $v_s \sim d_m \, k^c_+$ is the approximate velocity of slippage, and $\tau_c$ is a characteristic time. This characteristic time can be equivalent to the characteristic time of reptation of a polymer chain.  The velocity of slippage $v_s$ is approximately the multiplication of the length of the lattice site $d_m$ to the frequency a chain experiences slippage events after cleavage i.e.  $k^c_+$. Note that the form for $f$ in eqn (\ref{eq:force_der1}) is due to one unwound vulnerable site in presence of one enzyme. Due to multiple unwound vulnerable sites (in presence of enzymes), the total average force can be proportional to the rate at which enzymatic unwinding happens, i.e., $2 \, k^w_+ \, N^s_{E_V E_R}$ (see eqn (\ref{eq:kinetics4})). Thus the average force per remaining lattice sites can be 
\begin{equation}
    F \, \sim \, \frac{1}{N^s} \, \big(\gamma \, d_m \, k^c_{+} \, \tau_{c} \big) \, \, (2 \, k^w_+ \, N^s_{E_V E_R}) \, \tau_e ,
    \label{eq:force_der_2}
\end{equation}
where $\tau_{e}$ is  an average time required to form disentangled chains during detachment from the surface. When a fibril loses its lattice sites in degradation, dangling and entangled chains of the cleaved tropocollagen units (weakly attached to the fibril surface) appear continuously. In phenomenological manner and following the work of \cite{de1979scaling}, we propose $\tau_e \, = \, \tau^0_{r} + \tau_c \, \big({\rm exp} \, (\nu) - 1 \big)$ \citep{adjari1994slippage, vega2001terminal}, where $\nu$ is the number of lattice sites lost in form of disentangled chains during degradation.  When $\nu$ is zero, i.e. in absence of degradation,  $\tau_e = \tau_r^0$ can be treated as average relaxation time of a fibril. The value of $\tau_r^0$ is 7 s \citep{shen2011viscoelastic, gautieri2012viscoelastic}. We treated $\tau_c$  as  a material parameter and obtained its value  as 1.4 s using the  experimental data of \cite{flynn2013highly} single fibril degradation under zero loading condition (see section SI10 and Fig.~\ref{fig:res1_single_fibril}a). 

\subsection{Reaction rate constants and other parameters:} 

\noindent
We provided the values of rate constants and other parameters in  table~\ref{tab:table1}. Using the relation of protein adsorption-desorption rate constants \citep{adamczyk2000kinetics}, we set $k^1_+$ and $k^1_-$  where $k^1_- = k^1_+ \, {\rm exp} \, (-\Delta U_{ad}/(k_B T))$. 
We assumed that all unbinding events happen with same probability (backwards reactions of eqn (\ref{eq:kinetics1}) and (\ref{eq:kinetics2})), and the corresponding rate constants  are equal, i.e. $k^1_- = k^2_- = k^3_- = 5.65 \times 10^{-3}$ s$^{-1}$ \citep{ottl2000recognition}. Using the value of $k^1_-$ and for a chosen potential barrier difference of $\Delta U_{ad} \sim k_B T $, $k^1_+ = k^2_+ = 0.0154$ s$^{-1}$.  In experiments, the velocity of the enzyme on the collagen  surface reported was
 $\sim 4.5 \times 10^{-6}$ m/s \citep{saffarian2004interstitial}. Using this reported value of the velocity, we set the forward rate constants of eqn (\ref{eq:kinetics2}) and (\ref{eq:kinetics3}) related to  full-length binding/hopping. The minimum distance  an enzyme can traverse to find another lattice site for full-length binding or  jump to change track is $d_{\text{TC}} = 1.5$ nm, which sets $k^3_+ = k^4_+ = 3 \times 10^3$ s$^{-1}$. However, as the probability of the occurrence represented by the second part  of eqn (\ref{eq:kinetics3}) (see main text) is expected to be lower compared to the other hopping events, we set the value of $k^5_+$  in an approximate manner based on the minimum distance between two neighboring vulnerable sites which is $D_{gap} = $ 67 nm, and it sets $k^5_+ = (k^3_+/(D_{gap}/d_{\text{TC}}))$ s$^{-1}$. Using the transition state theory, we set the rate constant for enzyme-induced irreversible unwinding eqn (\ref{eq:kinetics4})  as  $k^w_+ = (k_B T/h) \; {\rm exp}\,(-2\, E_m/k_B T)$, where $E_m =  n_{res} \, \Delta G/N_A$, and  $(E_V E_R)^*$ corresponds to two lattice sites. For $\Delta G = 1.9$ kJ/mole per amino acid residue \citep{perumal2008collagen}, $n_{res} = 28$ for one lattice site, $N_A = 6.023 \times 10^{23}$ and  a chosen temperature $T = 310$ K, it yields to $k^w_+ = 7.5 \times 10^{-6}$ s$^{-1}$. The value of $k^c_+$ is different for different types of collagenase and collagen substrate \citep{welgus1982gelatinolytic, fields1987sequence, mallya1992kinetics, lauer2000hydrolysis,salsas2010cleavage}. To test the present model, we used two values: $k^c_+ = 0.472$ s$^{-1}$ (fibroblast collagenase and native collagen type I) \citep{welgus1982gelatinolytic} and $k^c_+ = 0.583$ s$^{-1}$ (class I clostridium histolyticum collagenases and rat type I) \citep{mallya1992kinetics}.

\begin{table}[hbtp]
    \centering
    \begin{tabular}{c c}
         \hline 
Parameter &  Value \\
\hline
$d_{\text TC}$ & 1.5 nm  \citep{chung2004collagenase} \\
$\ell_{\text TC}$ & 300 nm  \citep{chung2004collagenase} \\
D-band gap ($D_{gap}$) & 67 nm  \citep{chung2004collagenase} \\
$d_E$ & 10 nm   \citep{tyn1990prediction}\\
$d_m$ & 8 nm \citep{perumal2008collagen} \\
$n_{res}$ & 28 \citep{perumal2008collagen} \\
$c_1$ & 25  (discussed in main text) \\
 $\Delta G$ & 1.9 kJ/mole \\
 & per amino acid residue \\
 & \citep{perumal2008collagen} \\
 $\lambda_m$ & 3.6 \AA   
 \citep{perumal2008collagen} \\
 $k^1_+ = k^2_+$ & 0.0154 s$^{-1}$ \citep{adamczyk2000kinetics}\\
 $k^1_- = k^2_- = k^3_-$ & $5.65 \times 10^{-3}$ s$^{-1}$ \citep{ottl2000recognition}\\
 $k^3_+ = k^4_+$ & $3 \times 10^3$ s$^{-1}$  \citep{saffarian2004interstitial}\\
 $k^5_+$ & $(k^3_+/(D_{gap}/d_{\text{TC}}))$ \\
 $k^w_+$ & $7.5 \times 10^{-6}$ s$^{-1}$  \cite{perumal2008collagen} \\
 $k^c_+$ & 0.583 s$^{-1}$ \citep{welgus1982gelatinolytic}\\ 
 &  or, 0.472 s$^{-1}$  \citep{mallya1992kinetics}\\
  $\tau^0_r$ & 7 s \citep{shen2011viscoelastic} \\
  $\tau_c$ & 1.4 s (obtained using experiment \\
  &  of \cite{flynn2013highly}) \\
\hline
    \end{tabular}
    \caption{List of  parameters and their values.}
    \label{tab:table1}
\end{table}

\subsection{Initial conditions to solve the ODEs:}

\noindent
We used MATLAB ODE23 solver  to solve the ODEs using the following initial conditions. The nine first order ODEs (eqn (\ref{eq:enzyme})-(\ref{eq:product})) require nine following initial conditions: at time $t = 0$, number of enzymes surrounding a fibril $N^0_e$, $N^s_R = (N^s)^0 - (N^s_V)^0$, $N^s_V = (N^s_V)^0$, $N^s_{E_R E} = 0$, $N^s_{E_V E} = 0$, $N^s_{E_R E_R} = 0$, $N^s_{E_V E_R} = 0$, $N^s_{(E_V E_R)^*} = 0$, and $N_P = 0$. For a fibril of initial diameter $d_f (t = 0) = d_f^0$ and length $\ell_f (t = 0) = \ell_f^0$, we obtained $(N^s)^0$ and $(N^s_V)^0$  using eqn (\ref{eq:total_sites}) and  eqn (\ref{eq:tc_units}), respectively.

\subsection{Fibril fraction estimation in simulation box from collagen concentration:}

To set the fibril fraction $\phi_f$ (the ratio of the volume of all collagen fibrils to the volume of the simulation box) in our simulations, we  followed a crude analytical way  converting the collagen concentration to $\phi_f$. An example is the following. We consider a collagen concentration $\sim$ 2.5 mg/mL \citep{ranamukhaarachchi2019macromolecular}. 
Note that we  performed {\it in vitro} experiments using the same collagen concentration.
For this chosen concentration and using the molecular weight of tropocollagen  $\sim 300$ kDa \citep{leon2019hydrolyzed}, the  number of tropocollagen units in the simulation box volume (which is 125 \textmu m$^3$) is $\sim 0.63 \times 10^6$. The volume of one tropocollagen  is $\frac{\pi}{4} \, (d_{\text TC})^2 \, \ell_{\text TC}$. Thus the total volume of all tropocollagen molecules in the simulation box volume  is  $334 \times 10^{-3}$ \textmu m$^3$. 

These tropocollagen molecules arrange and generate  fibrils  in a solvent medium. Because of the gaps in between tropocollagen molecules in a fibril in hydrated condition (which is difficult to estimate in the present work), the total volume of all fibrils must be higher than  the total volume of all tropocollagen molecules. To obtain a nearly  correct estimate of the total volume of all fibrils, we assumed packing factor for cylindrical tropocollagen in loosely packed condition as 0.4 (note that the value of  packing factor for spheres in loosely packed state is in the range $0.4-0.56$). Thus the nearly corrected volume of all fibrils is $(334 \times 10^{-3}$ \textmu m$^3)/0.4$, and this volume yields to a fibril fraction $\phi_f \sim 0.006$ for the chosen  collagen concentration 2.5 mg/mL . 

The collagen percentage in different organs (such as, brain, liver, heart, kidney, lung and colon, etc.) varies in the range $0.1-6$\% \citep{tarnutzer2023collagen}.  We  performed simulations with different values of $\phi_f$ in the range $0.003-0.03$, which is  $0.3-3$\%.

\subsection{Experimental methods:}

{\textbf{Chemicals:}} 

\noindent
High concentration, rat tail acid extracted type I collagen was procured from Corning (Corning, NY). PEG (8000 Da) was ordered in powder form from Sigma-Aldrich (St Louis, MO) and reconstituted in PBS (Life Technologies, Carlsbad, CA) immediately before usage with a final concentration of 100 mg/mL, and 1× reconstitution buffer was composed of sodium bicarbonate, HEPES free acid, and nanopure water. 

\vspace{0.3cm}
\noindent
{\textbf{Preparation of collagen gels with different microstructures:}} 

\noindent
First, PEG of required amounts to make a 2 or 8 mg/mL final concentration (denoted as P2 or P8) was added to the DMEM. This is followed by addition of the reconstitution buffer and mixing. Thereafter, the collagen stock was added to the mixture to produce a final concentration of 2.5 mg/mL. Finally, pH of the final mixture was adjusted using 1 N NaOH, followed by incubation ($\sim$ 45 minutes) at 37$^{\text o}$C. Following polymerization, PEG was washed out of the gels by rinsing with the DMEM (3× for 5 minutes each). For collagenase treatment, gels were incubated with bacterial collagenase of 2.5 \textmu g/mL for about 45 minutes. (We added 50 microliters of 10 microgram per mL of collagenase on top of the gels which yields final concentration of collagen to be 2.5 \textmu g/mL.)

\vspace{0.3cm}
\noindent
{\textbf{Fast green staining and imaging of collagen gels:}} 

\noindent
The prepared collagen gels were fixed with 4\% paraformaldehyde for 30 mins. After that, the gels were washed thoroughly at least 3 times by subjecting them to shaking and replacing with PBS for 10 min. The gels were then incubated with 100 \textmu g/mL of fast green dye in PBS (Fast green FCF, Thermo Fischer, USA) and were subjected to shaking overnight. The gels were  washed with PBS at least 3 times in an orbital shaker for a duration of 30 minutes each time. The stained gels were then imaged using a confocal fluorescence microscope (Leica, SP8) with 40x water immersion objective. Fast green was excited at a wavelength of 627 nm, and 630-730 nm was used for detection. 

\vspace{0.3cm}
\noindent
{\textbf{Scanning electron microscopy (SEM):}} 

\noindent
SEM was performed on the gels using FEI SEM Apreo equipped with ETD detector. Both the collagenase treated gels and the non-treated gels were first fixed with 4\% paraformaldehyde for 1 hour. This is followed by 3x rinsing in PBS for 10 minutes in each step of shaking. The gels were then rinsed twice with Milli-Q water for 15 minutes each. The fixed gels were then subjected to a series of dehydration steps in ethanol and hexamethyldisilazane (HMDS) using the protocol outlined in \cite{raub2007noninvasive}. Briefly, the gels were dehydrated first in ethanol dilution series: 30\%, 50\%, 70\%, 90\%  and 100\% for 15 minutes of each step. The gels were then incubated in ethanol/HMDS dilution series: 33\%, 50\%, 66\% and 100\% for 15 minutes each. After the final incubation, the gels were allowed to dry on an aluminum foil for at least 1 day in the fume hood. The dried gel samples were then sputter coated with a Pelco SC-7 sputter coater with gold as the target. The gels were then imaged at 5 kV and 0.6 nA with magnifications of 15000x and 30000x.

\subsection{Single fibril model validation:}

\noindent
We  used the experimental data of \cite{flynn2013highly} on single fibril degradation   to calibrate and validate our single fibril model. \cite{flynn2013highly} reported the diameter of degrading single collagen fibril under different external loading condition in a 5 \textmu M Clostridium histolyticum bacterial collagenase type A solution. To solve our single fibril model for the parameter sets reported in \cite{flynn2013highly}, we need to obtain the number of enzymes $N_e^0$ for a fibril. Using their enzyme concentration,  we  estimated $N_e^0$ for a fibril  in an approximate manner based on a volume element  equivalent to the initial volume of the fibril. For fibrils of  initial diameter $d_f^0 = 415$ nm and  $d_f^0 = 250$ nm, the values of $N_e^0$ are  2100 and 750, respectively, for a chosen length $\ell_f^0 = 4000$ nm.  The value of $\tau_c = 1.4$ s is obtained by calibrating our model prediction to the single fibril data corresponding to  $d_f^0 = 415$ nm under zero load condition (see Fig.~\ref{fig:res_single_fibril}a). The value of $k^c_+$ used to calibrate our model is  0.583 s$^{-1}$ (class I clostridium histolyticum collagenases and rat type I) \citep{mallya1992kinetics}.  In a nutshell, our single fibril model have satisfactorily captured the  experimental findings of \cite{flynn2013highly} (see Fig.~\ref{fig:res_single_fibril}a).

\vspace{0.2cm}
\noindent
\textbf{The choice of} $\bm{\delta E_m = 0.013 \, E_m}$:  The degradation rate reduces when the fibril is under external tension \citep{bhole2009mechanical, camp2011molecular,  flynn2013highly} (see Fig.~\ref{fig:res_single_fibril}a). Perhaps the external tension increases the stability of the triple helices in the fibril by increasing the energy barrier for enzymatic unwinding \citep{chang2012molecular, chang2014molecular, tonge2015micromechanical, saini2020tension, topol2021fibrillar}. However, the cleavage mechanism can be different for the isolated tropocollagen (triple helix) molecules under external tension \citep{adhikari2011mechanical, adhikari2012conformational}. For a fibril, the increase in the internal energy due to external loading can increase the energy barrier for enzymatic unwinding where the rate is proportional to $k^w_+ \propto {\rm exp}\,(-2 E_m/k_B T)$. The increase in the energy barrier results in the decrease in the rate of unwinding, implying a decrease in the rate of degradation.

It is difficult to provide a direct estimation of the increase in the energy barrier of enzymatic unwinding because of external loading. We have followed a simple analytical approach to obtain the increment in the energy following the work of \cite{tonge2015micromechanical} and using a general expression for strain energy (purely elastic). 
Under external loading, we can write the increment in energy $\Delta U$ for a tropocollagen under low external load  $f_{ext}=$ 2 pN is
\begin{equation}
\Delta U \sim \frac{1}{2} \, \Big( \frac{\pi}{4} \, d^2_{\text TC} \,  (\ell_{\text TC} + \delta \ell) \Big) \, \, \frac{f_{ext}}{(\pi/4) \, d^2_{\text TC}} \, \, \frac{f_{ext}}{(\pi/4) \, d^2_{\text TC} \, E_{\text TC}},
\label{eq:strain_en}
\end{equation}
where $\delta \ell$ is the extension  and $E_{\text TC}$ is the elastic modulus of a tropocollagen. The value of $E_{\text TC}$ is in the range $0.01-1$ GPa  \citep{david1978biomechanics, sun2002direct, 
gautieri2012viscoelastic}. For a chosen value $E_{\text TC} \sim 0.5$ GPa and $\delta \ell$ in the range $ 50 - 250$ nm \citep{sun2002direct}, (\ref{eq:strain_en}) yields to $\Delta U \sim 0.009 \, E_m - 0.015 \, E_m$, where $E_m =  n_{res} \, \Delta G/N_A$. Our model captured the experimental trend \citep{flynn2013highly} well for $\delta E_m = 0.013 \, E_m$.

\begin{figure}[hbtp]
\includegraphics[width=\textwidth]{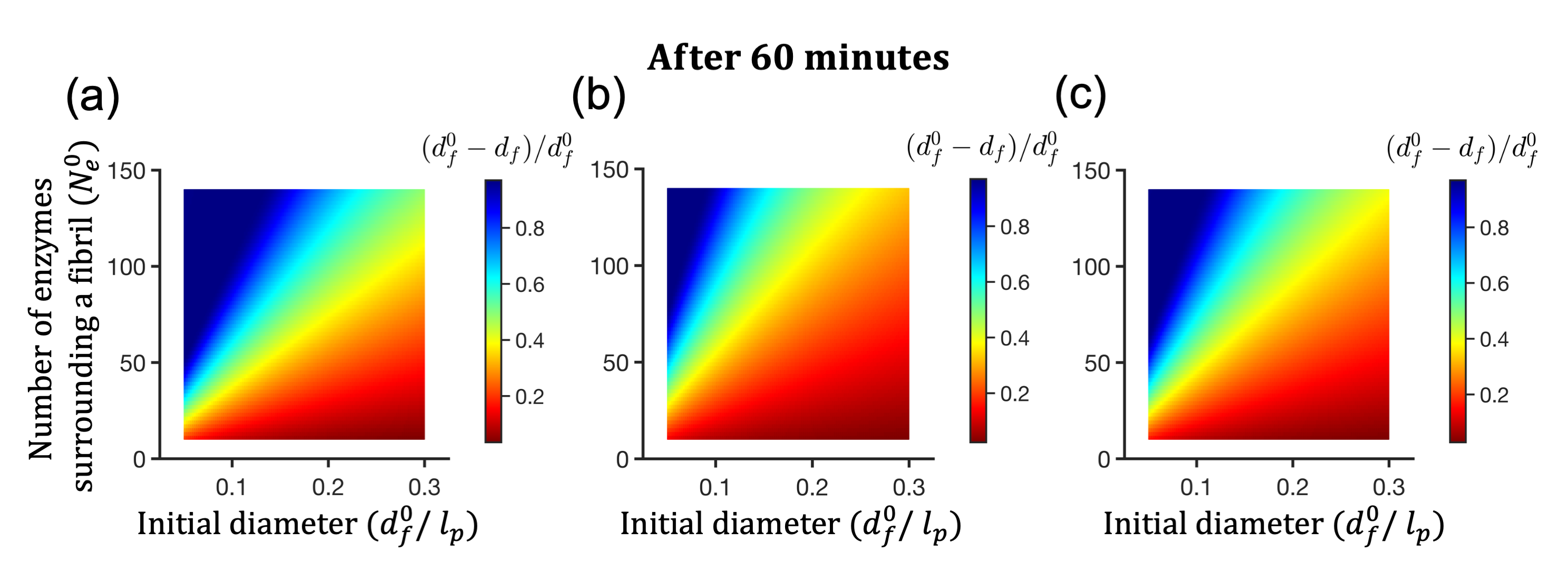}
\caption{{\textbf{Degradation of single collagen fibril.}} (a)-(c) are color maps representing the extent of degradation after 1 hour for ranges of $N_e^0$ (number of enzymes surrounding a fibril) and $d_f^0/l_p$ (initial fibril diameter) where $l_p = 1\,\mu$m is a nominal length scale. (a) fibril length $\ell_f/l_p = 2$,  $k^c_+ = $ 0.583 s$^{-1}$ \citep{mallya1992kinetics}; (b) fibril length $\ell_f/l_p = 3$, $k^c_+ = $ 0.583 s$^{-1}$; (c) fibril length $\ell_f/l_p = 2$, $k^c_+ = $ 0.472 s$^{-1}$ \citep{welgus1982gelatinolytic}. } 
\label{fig:res1_single_fibril}
\end{figure}

\vspace{0.2cm}
\noindent
\textbf{The scaling related to single fibril degrdation}: Figs.~\ref{fig:res_single_fibril}b,c imply that the degradability must be directly proportional to the  number of enzymes per unit surface area of the fibril. According to the model assumptions, if the enzymes are not loosing their activity and potency as enzymatic degradation progresses,  then  the surface area of the fibril is the only parameter which decreases with time. Thus a thicker fibril and a thinner fibril of same length can degrade up to a same extent if the ratio $N_e^0/A^0_f$, where $A^0_f = (\pi d_f^0 \ell_f)$ is the area of the fibril, is fixed. This is reflected in Fig.~\ref{fig:res1_single_fibril}a-c, where the same extent of degradation is observed if $N_e^0/d_f^0$ is a constant. The model predicts thicker fibrils with significantly larger number of enzymes to be highly degradable than thinner fibrils with very less number of enzymes subject to the ratio $N_e^0/d_f^0$, and \textit{vice-versa}.  The predictions are sensitive to the choice of $k^c_+$ which varies with different MMPs and types of collagen. As expected, the extent of degradation decreases with the decrease in $k^c_+$. 

\subsection{Model microstructure generation for P2 and P8 gels   using the histograms of fibril diameters from the experiments:}

\begin{figure}[hbtp]
\includegraphics[width=1\linewidth]{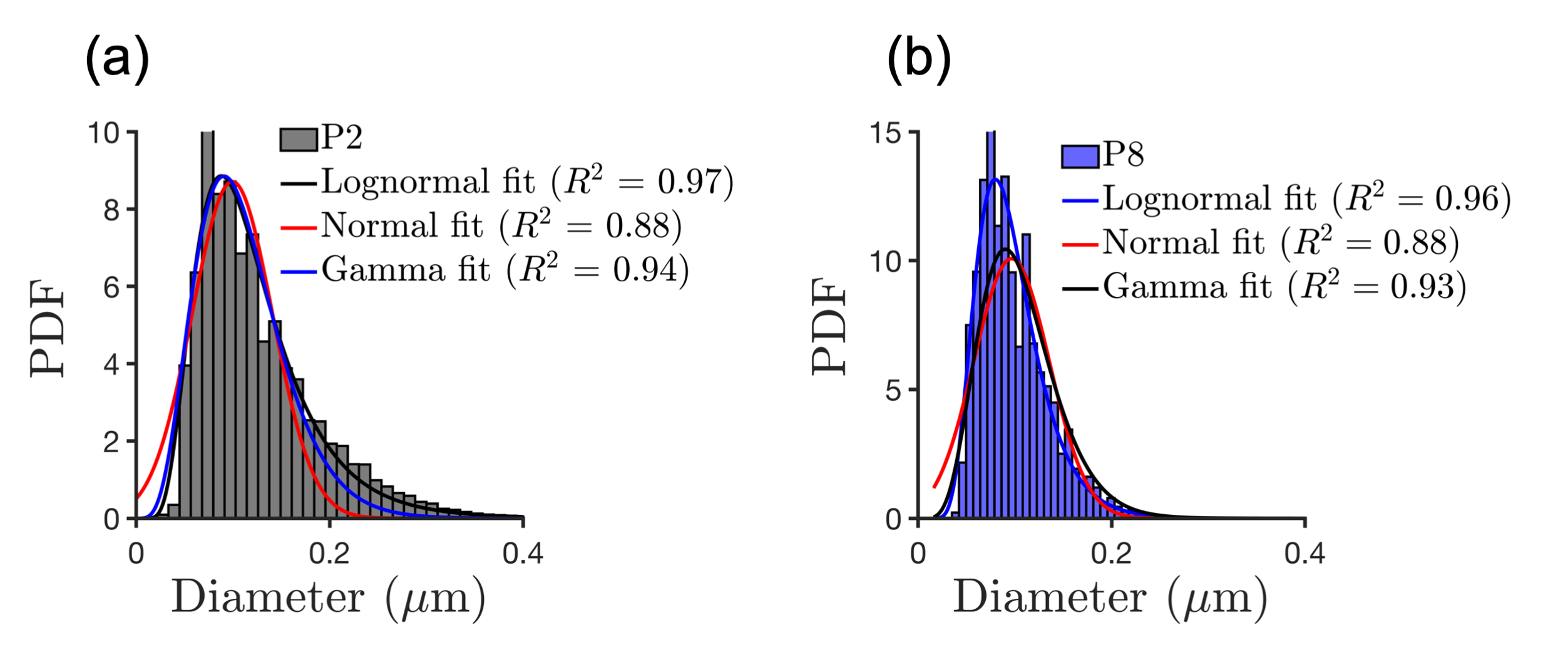}
\caption{Different fits and goodness of fits for P2 (a) and P8 (b).}
 \label{fig:goodness_fits}
\end{figure}

\noindent
The lognormal distribution curves fit our histrograms better than other distributions such as normal and gamma distributions (see Fig.~\ref{fig:goodness_fits}). 

\noindent

\begin{figure}[hbtp]
\includegraphics[width=1\linewidth]{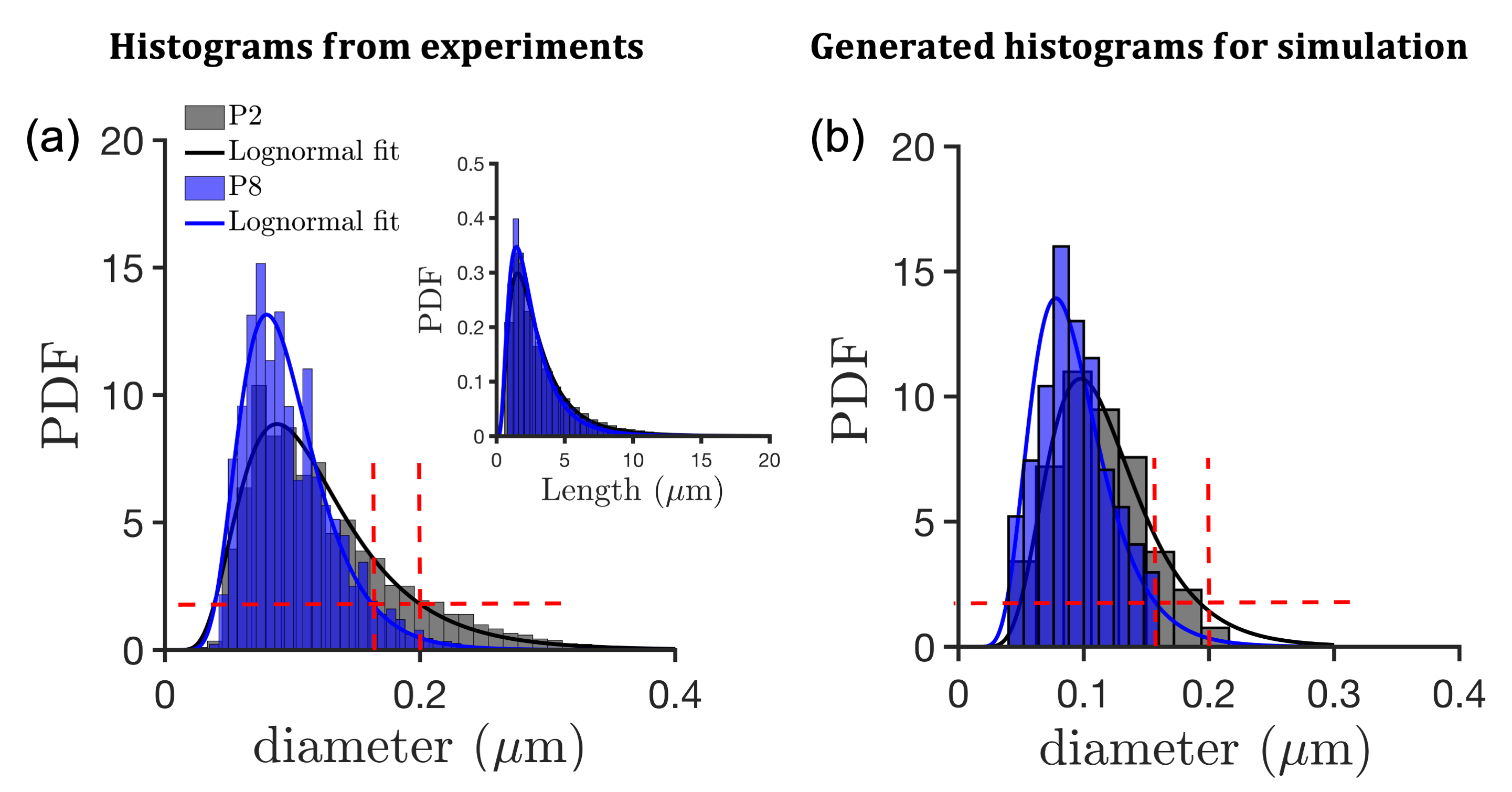}
\caption{Histograms of diameters for P2 and P8 matrices. (a) From experiments. (b) Histograms are generated for hybrid simulation using (a).}
 \label{fig:matrix_gen}
\end{figure}

We  used the lognormal fitted curves of the diameter histograms (P2 and P8) from the experiments  to generate the fibril distributions. The parameters related to the lognormal fits from P2 and P8 are $\mu_{P2} = -2.247$ and $\sigma_{P2} = 0.45$, and $\mu_{P8} = -2.288$ and $\sigma_{P8} = 0.4$, respectively. Using these values and lognormal function, random numbers are generated in between 0.04-0.2 \textmu m for P2,  0.03-0.16 \textmu m for P8, and we ignore the tail regions where PDF is less than 2 (see Fig.~\ref{fig:matrix_gen}a). In both P2 and P8, all fibrils are of same length. For P2, $\ell_f/l_p = 2$, and $\ell_f/l_p = 1.8$ for P8 which is 10\% smaller than that of P2. We note in passing that in experiments, the mean length of P8 is found to be slightly (approx. 10 \%) smaller than that of P2. For both P2 and P8, the distributions are generated such that the volume fraction of fibrils $\phi_f = 0.007$ turns out to be the same. The outcomes of number of fibrils $n_f$ in P2 and P8 for simulation are 40 and 75, respectively. The chosen value of $\phi_f$ is close to the collagen concentration used in the experiments 2.5 mg/mL.   For each P2 and P8, we performed three independent simulations, and we generated all three sets   using the histograms of diameters from experiments (Fig.~\ref{fig:matrix_gen}a). We showed the thickness distributions of the generated fibrils  in Fig.~\ref{fig:matrix_gen}b. The distributions generated for the simulations are qualitatively similar  to those from the experiments.

\noindent

\end{document}